# Cryogenic in-memory computing using magnetic topological insulators

Yuting Liu[1,2*], Albert Lee[3*], Kun Qian[1,4*], Peng Zhang[3], Zhihua Xiao[1, 5], Haoran He[3], Zheyu Ren[1,4], Shun Kong Cheung[1], Ruizi Liu[1,4], Yaoyin Li[2], Xu Zhang[1], Zichao Ma[1], Jianyuan Zhao[2], Weiwei Zhao[2], Guoqiang Yu[6], Xin Wang[7], Junwei Liu[4, 8], Zhongrui Wang[9], Kang L. Wang[3], & Qiming Shao[1,4,5,8,10†]

[1]Department of Electronic and Computer Engineering, The Hong Kong University of Science and Technology, Clear Water Bay, Kowloon, Hong Kong SAR 999077, China

[2]School of Integrated Circuit, Harbin Institute of Technology, Shenzhen 518055, China.

[3]Device Research Laboratory, Department of Electrical and Computer Engineering, University of California, Los Angeles, California 90095, USA.

[4]IAS Center for Quantum Technologies, The Hong Kong University of Science and Technology, Hong Kong, China

[5]ACCESS – AI Chip Center for Emerging Smart Systems, InnoHK Centers, Hong Kong Science Park, Hong Kong, China

[6]Beijing National Laboratory for Condensed Matter Physics, Institute of Physics, University of Chinese Academy of Sciences, Chinese Academy of Sciences, Beijing, 100190 China

[7]Department of Physics, The City University of Hong Kong, Hong Kong SAR 999077, China

[8]Department of Physics, The Hong Kong University of Science and Technology, Clear Water Bay, Kowloon, Hong Kong SAR 999077, China

[9]Department of Electrical and Electronic Engineering, the University of Hong Kong, Pokfulam Road, Hong Kong SAR 999077, China

[10]Guangdong-Hong Kong-Macao Joint Laboratory for Intelligent Micro-Nano Optoelectronic Technology, The Hong Kong University of Science and Technology, Hong Kong, China

*Equal contribution

†Email: eeqshao@ust.hk

**Machine learning algorithms have been proven effective for essential quantum computation tasks such as quantum error correction and quantum control. Efficient hardware implementation of these algorithms at cryogenic temperatures is essential. Here, we utilize magnetic topological insulators as memristors (termed magnetic topological memristors) and introduce a cryogenic in-memory computing scheme based on the coexistence of the chiral edge state and the topological surface state. The memristive switching and reading of the giant anomalous Hall effect exhibit high energy efficiency, high stability, and low stochasticity. We achieve high accuracy in a proof-of-concept classification task using four magnetic topological memristors. Furthermore, our algorithm-level and circuit-level simulations of large-scale neural networks demonstrate software-level accuracy and lower energy consumption for image recognition and quantum state preparation compared with existing magnetic memristor and CMOS technologies. Our results not only showcase a new application of chiral edge states but also may inspire further topological quantum physics-based novel computing schemes.**

Quantum bit (qubit) can be abstracted into a two-level system, for which the physical implementation can be based on superconducting circuits, semiconductor dots, ion traps, optical photons, and others[1]. An alternative promising solution is topological qubits that can be constructed by hybridizing chiral edge states and superconducting orders[2]. Qubits can be controlled and read out by tailored short pulses so that processing quantum information is possible. At this moment, tens of superconducting qubits can be integrated into a single chip to demonstrate quantum advantages, which already requires 205 microwave cable[3]. As the number of controllable qubits increases, the number of input-output ports will increase dramatically, which demands a clear plan for scalability. Inspired by the complementary metal-oxide-semiconductors (CMOS) technology that integrates billions of transistors, multiple inputs and outputs can be combined using multiplexers and demultiplexers to reduce the number of input/output ports. This kind of peripheral circuit needs to sit beside the quantum chip and thus works at deep cryogenic temperatures. Cryogenic CMOS using von Neumann architecture works well for traditional tasks at this temperature[4–7]. As a result, cryogenic electronics have become essential in reducing the number of input/output ports to the quantum chips and generating multiplexed reading and control pulses for scalable quantum computation[4–8]. However, when it handles machine learning algorithms to perform quantum error correction[9] and quantum control[10], its performance and efficiency are limited by physically separated memory and processing units, the so-called von Neumann bottleneck, which incurs large time and energy overheads. A demanding request for cryogenic electronics is to support efficient machine learning algorithms. To address this challenge, bio-inspired computing architectures with co-location of memory and processing units such as in-memory computing have been proposed. In-memory computing using memristors, which are nonvolatile and electrically programmable devices, eliminates the huge amount of energy-costly and slow data transfer between computation and memory units, promising energy-efficient hardware implementation of machine learning algorithms[11–13]. These memristors act as artificial synapses in neural networks and their crossbar arrays physically embody weight matrices. Such a design allows them to compute entire matrix-vector multiplications in a single cycle, which is the essential computation step for artificial neural networks used in deep learning and overcomes the von Neumann bottleneck in traditional computing architectures[14].

Cryogenic in-memory computing for quantum computation requires energy-efficient memristors working at deep cryogenic temperatures (liquid helium temperature 4.2 K or below), which remain elusive. Several implementations of memristive crossbars have been developed based on different devices at room temperature, including (redox- or conductive bridge-based) resistive devices[15,16], phase change devices[17], ferroelectric devices[18,19], and magnetic devices[20]. The cryogenic memristor array remains to be experimentally explored. Magnetic devices can work at cryogenic temperatures (liquid nitrogen temperature 77 K or below) as a binary and nonvolatile memory[21,22]. However, the energy efficiency of the cryogenic magnetic device is comparable with its room temperature counterpart since the spin current generation is based on conventional spin-transfer torque or spin Hall effect, which is rather temperature-insensitive. Due to the limited cooling power at deep cryogenic temperatures[4], a more energy-efficient analog magnetic memristor is needed. Magnetic topological insulators are promising candidates due to their tunable magnetic order by electrical currents with high energy efficiency[23–28]. In this work, we introduce chiral edge state-based magnetic topological memristors (MTMs) by using magnetic topological insulator Hall bar devices. On the one hand, the chiral edge state exhibits giant and bipolar anomalous Hall resistance, which facilitates the electrical readout. On the other hand, the magnetic order and thus the anomalous Hall resistance can be tuned through spin-momentum locked topological surface current injection. We demonstrate the analog memristive switching behavior in MTMs and a proof-of-concept classification system using four MTMs. The algorithm-level and circuit-level simulations of hybrid MTM-CMOS-based neural networks indicate a software-level accuracy and lower energy consumption compared with existing memristor technologies.

**Memristive behaviors in MTIs**

Our memristor is based on a magnetic topological insulator (MTI). We prepare MTIs, Cr-doped $(Cr_{0.15}Bi_{0.26}Sb_{0.59})_2Te_3$ (Cr-BST), using molecular beam epitaxy and fabricate them into Hall bar devices (see Methods 'Device fabrication and characterization' and **Extended Data Figs. 1 and 2**). Due to topologically nontrivial band structures, these MTIs host chiral edge states (**Fig. 1a**). At zero temperatures, the chiral edge state is dissipationless, and anomalous Hall resistance is quantized to $h/e^2$, where $e$ is the electron charge and $h$ is Planck's constant. At finite temperatures, dissipative surface states appear, and the chiral edge state becomes dissipative arisen from effective scattering between two edges due to the presence of bulk or surface states [28,29]. Nevertheless, as long as the contribution of the chiral edge states remains significant in electronic transport, MTIs can exhibit a giant anomalous Hall resistance. In our experiment at 2 K, the saturated anomalous Hall resistance for devices D1-D4 is 11 kΩ (0.42 $h/e^2$) for an excitation current of 83 A/cm$^2$ (**Fig. 1b**). The tangent of anomalous Hall angle that characterizes the ratio of transverse resistance over longitudinal resistance can reach 0.6 (see Methods 'Co-existence of CES and TSS in MTI' and **Extended Data Figs. 3 and 4**), significantly larger than those of the topologically trivial magnetic materials, indicating a large contribution from the chiral edge state[27,28]. In another device D5 with the nominally same growth recipe, we achieve the tangent of 2.67, indicating an even larger contribution from the chiral edge state (see Methods). Since the MTI at 2 K is not in the quantum anomalous Hall insulator state (as device D5 shows at 100 mK in **Extended Data Fig. 3c**), there is still finite contribution from the topological surface state (**Fig. 1c**), which is also evidenced from the large non-reciprocal magnetoresistance (see Supplementary Note 1). Thanks to the spin-momentum locking of topological surface states, the spin-polarized topological surface current has been demonstrated to generate giant spin-orbit torque (SOT) and manipulate the magnetic order of MTIs efficiently[23–25]. We measured a SOT efficiency $\xi_{DL}$ of 19.2, which is much larger than the control sample Ta and reported values from heavy metals (see Supplementary Note 2 and Supplementary Table 1). More importantly, the SOT from the spin-momentum locking is orders of magnitude larger at cryogenic temperature than that at room temperature[26], which is different from spin-transfer torque or spin Hall effect-induced SOT. With this giant SOT, the theoretical normalized switching power that is proportional to $\rho/\xi_{DL}^2$ is much lower for MTIs compared with heavy metal cases, where $\rho$ is channel resistivity (see Supplementary Note 2 and Supplementary Table 1). We apply a series of pulsed charge currents into the MTI Hall bar device and then measure the corresponding Hall resistance (**Fig. 1c**). **Fig. 1d** shows the current-induced magnetization switching in four MTI Hall bar devices D1-D4 through the SOT effect. The anomalous Hall resistance is tunable and ranges between -600 Ω and 600 Ω. The large reduction of the current switching range from the field switching range is attributed to the Joule heating effect, which breaks down the MTI into a multi-domain state[30] (see Supplementary Note 3). Nevertheless, the large anomalous Hall resistance of about 600 Ω is still very large, indicating that chiral edge states play an important role in transport. The experimental switching current density $J_{sw}$ ($4.2 \times 10^5$ A/cm$^2$) and corresponding normalized switching power that is proportional to $\rho J_{sw}^2$ are much lower for MTIs compared with heavy metal cases (see Supplementary Note 2 and Supplementary Table 1). The basic properties of all mentioned devices are summarized in Supplementary Table 2.

To utilize MTI as a memristor, we need to characterize its write and read capability from an application point of view. For better memristor-based technology, the number of available states for an MTM should be as large as possible and these states should be stable. We build a platform to experimentally test multiple devices at cryogenic temperature (**Fig. 2a**), which allows us to apply arbitrary pulse sequences. We have 50 trials of write tests for 12 different levels and they exhibit very low write stochasticity of 1.9% (**Fig. 2b**) (see Supplementary Note 4), which is significantly lower than other nonvolatile memory technologies and beneficial for neural network implementation[12]. The current density used to reset the magnetization state can be as low as $7\times10^5$ A/cm$^2$, which indicates the high efficiency of spin-orbit torque and is consistent with the previous reports[23–25]. Also, the switching is almost analog and thus the number of available states is much larger than 12 (**Fig. 1d**). We have 90 trials for reading tests for the same 12 levels by using the fixed read pulse magnitude and they exhibit even lower read stochasticity of 0.37% (**Fig. 2c**) (see Supplementary Note 4). The high energy efficiency and low stochasticity of write and read suggest that the MTI is a good choice for memristors. We also investigate the scalability of the analog switching behaviors and show that we can obtain at least 15 distinguishable states for a 120 nm × 200 nm MTI device (see Methods 'Scalability and multi-states of MTI memristors' and **Extended Data Fig. 5**).

**MTI array for data classification**

A crossbar of memristors leverages Ohm's law and Kirchhoff's current law to achieve analog multiply-accumulate operation, which is part of vector-matrix multiplication and the most frequent operation for neural network-based deep learning[31]. We demonstrate a proof-of-concept experiment of chiral edge state-based cryogenic in-memory computing by classifying the type of Iris flowers using four MTMs. In our experiment, the input is encoded in the input current and the output is encoded in the anomalous Hall voltage, where the anomalous Hall resistance is the weight of the matrix in the vector-matrix multiplication (**Fig. 2d**) (see Methods 'Implementation of Iris pattern classification'). We utilize three single-layer perceptrons and a softmax to classify three types of Iris flowers. We obtain the 12 software-trained weight parameters and then apply the corresponding currents to tune the anomalous Hall resistance. For each perceptron, we experimentally determine the classification accuracy 30 times. **Fig. 2e** shows that the accuracy fluctuates around 88% and can reach 96%, matching the software level accuracy. The variation in accuracy across trials is due to the read current-induced thermal noise disturbance (see Methods 'Implementation of Iris pattern classification' and **Extended Data Fig. 6**).

To show the feasibility of using MTMs for practical deep learning, we use the extracted device properties to perform neural network simulation tasks at a larger scale, including Modified National Institute of Standards and Technology (MNIST) handwritten digit recognition, Canadian Institute for Advanced Research (CIFAR-10) image recognition, and reinforcement learning for quantum state preparation. One unique feature of a chiral edge state-based memristor is its straightforward representation of both positive and negative weights, which is not available for traditional resistance-based memristors. The necessary condition of this anomalous Hall effect-based in-memory computing is that the anomalous Hall resistance needs to be sufficiently large to be read out effectively (see Methods 'Design challenges for Hall effect-based neural network').

For MNIST, a multi-layer perceptron is used (**Fig. 3a**), and for CIFAR-10, a convolutional neural network is used. We compare the performance of three neural networks built upon different weight constraints: bipolar weights (e.g., the MTM), unipolar weights (e.g., traditional resistance-based memristors), and floating-point weights (e.g., software; see Supplementary Note 5). We observe a similar performance between floating-point weights and MTM neural network, whereas the unipolar weight neural network exhibits significantly lower accuracy (**Fig. 3b** for MNIST and **Extended Data Fig. 7** for CIFAR) even with the adoption of algorithms to improve its performance[32]. The final normalized weight matrices of three neural networks are visualized in **Fig. 3c**. MTM and floating-point neural networks exhibit similar weight matrix patterns. In contrast, the limitation of positive weights in the unipolar weight neural network is insufficient to achieve the optimized weight matrix.

### MTI neural network for quantum control

To show the relevance of cryogenic in-memory computing for quantum computing, we then investigate the performance of MTM-based reinforcement learning for quantum state preparation (see Methods 'Qubit preparation with policy gradient' for details). The task aims to control the state of N serially coupled spins via a magnetic flux pulse sequence and drive it from an initial state to a target state (**Fig. 3d**). A policy gradient learning environment is prepared, and we again compare the neural networks with the aforementioned weight constraints. The MTM network is on par with the floating-point network (**Fig. 3e**). In contrast, the traditional memristor network performs worse due to limitations in its weight representation. In terms of the training time cost, the MTM network also outperforms the unipolar memristor network and achieves similar performance to the floating-point network (**Fig. 3f**).

### Circuit simulation of MTI neural networks

High-quality MTIs have been grown at wafer-scale on both crystalline and amorphous substrates using molecular beam epitaxy[33] and magnetron sputtering[34], making MTI a scalable and CMOS-compatible material system for cryogenic in-memory computing. We consider the task of designing the scalable circuit and system for MTM-based in-memory computing. This task is highly nontrivial since there has yet to be a hardware realization of anomalous Hall resistance-neural networks. Prior efforts have overlooked that the chiral edge state-based memristor is essentially a four-terminal device, where the sneak path doesn't allow for the simple parallel or series summation of two-terminal memristors[12,31] (see Methods "Design challenges for Hall effect-based neural network' and **Extended Data Fig. 8**). In contrast to previous works that use the summation of anomalous Hall voltage, we propose to leverage the summation of anomalous Hall current to perform matrix-vector multiplication. We have experimentally shown that the anomalous Hall current is proportional to both the applied longitudinal voltage and the z-direction magnetization (See **Extended Data Fig. 9**). Then, we verify the anomalous Hall current summation by connecting Hall current in series (See **Extended Data Fig. 10**). Based on this proposal, we design a hybrid MTM-CMOS system to realize the in-memory computing functionalities and successfully demonstrate the functionalities using a foundry-provided CMOS process design kit (See Supplementary Notes 6 and 7 for details on 'MTM neural network design' and 'Circuit simulation'). The proposed circuit implementation has shown a significant advantage in energy efficiency compared with CMOS technology only and the MRAM-based approach (see Supplementary Tables 4 and 5).

### Discussions

We also consider the application of our proposal using other promising material systems, such as intrinsic antiferromagnetic insulators and Moiré heterostructures such as $MnBi_2Te_4$[35] and twisted bilayer graphene[36,37]. These material systems can exhibit a quantized anomalous Hall effect of 25.8 kΩ ($h/e^2$) like our MTI[33,38] even at a higher temperature, which is beneficial for easy readout. In particular, current-induced magnetization switching with an ultralow current density[36,37] and electric field control of magnetic order[39] have been demonstrated in twisted bilayer graphene systems. To apply these systems for cryogenic in-memory computing, a systematic study of their memristive behaviors and scalable methods of fabricating a device array need to be studied like this work.

In summary, our work serves as a proof-of-concept demonstration of a cryogenic in-memory computing scheme based on chiral edge states. In light of the dissipationless nature of chiral edge states in the quantum anomalous Hall insulator state, pushing our MTI device to the quantized Hall resistance regime can be potentially beneficial[40]. However, the absence of the topological surface state may lead to the diminishment of the SOT, calling for an optimization of the contributions from different states. Nevertheless, very recently, the current-induced magnetization switching in the quantum anomalous Hall state with the assistance of the heating effect and gate voltage tuning was demonstrated[41]. Besides, there is a large family of quantum material systems that host chiral edge states and other tunable collective orders. We envision that many of these material systems can be utilized for cryogenic in-memory computing.

**Acknowledgements**

We thank B. Lian and X. Sun for fruitful discussions. The authors at HKUST acknowledge funding support from National Key R&D Program of China (Grants No. 2021YFA1401500), NSFC/RGC Joint Research Scheme (No. N_HKUST620/21), Shenzhen-Hong Kong-Macau Science and Technology Program (Category C) (SGDX2020110309460000), Research Grant Council-Early Career Scheme (Grant No. 26200520), HKUST-Kaisa Joint Research Institute grant (NoOKT21EG08) and Research Fund of Guangdong-Hong Kong-Macao Joint Laboratory for Intelligent Micro-Nano Optoelectronic Technology (Grant No. 2020B1212030010). This research was partially supported by ACCESS - AI Chip Center for Emerging Smart Systems, sponsored by InnoHK funding, Hong Kong SAR, and the State Key Laboratory of Advanced Displays and Optoelectronics Technologies. Y. Liu acknowledges the funding support from the HKUST Postdoc Fellowship Matching fund (NA389), the Harbin institute of technology (Shenzhen) startup funding for high talents, and the NSFC youth program (Grant No. 12304137).

**Author Contributions Statement**

Q.S. and K.L.W. conceived the experiments. P.Z. grew films. Y.L., Z.R., R. L., X.Z. and Z.M. fabricated the device. Y.L., K.Q. and S.C. conducted the electrical measurements with help from J. Z., A.L., Z.W. and H.H. performed the simulation. Y.L., A.L. and Q.S. drafted the manuscript and all authors reviewed the manuscript.

**Competing Interests Statement**

The authors declare that they have no competing financial interests.

**Figures and captions**

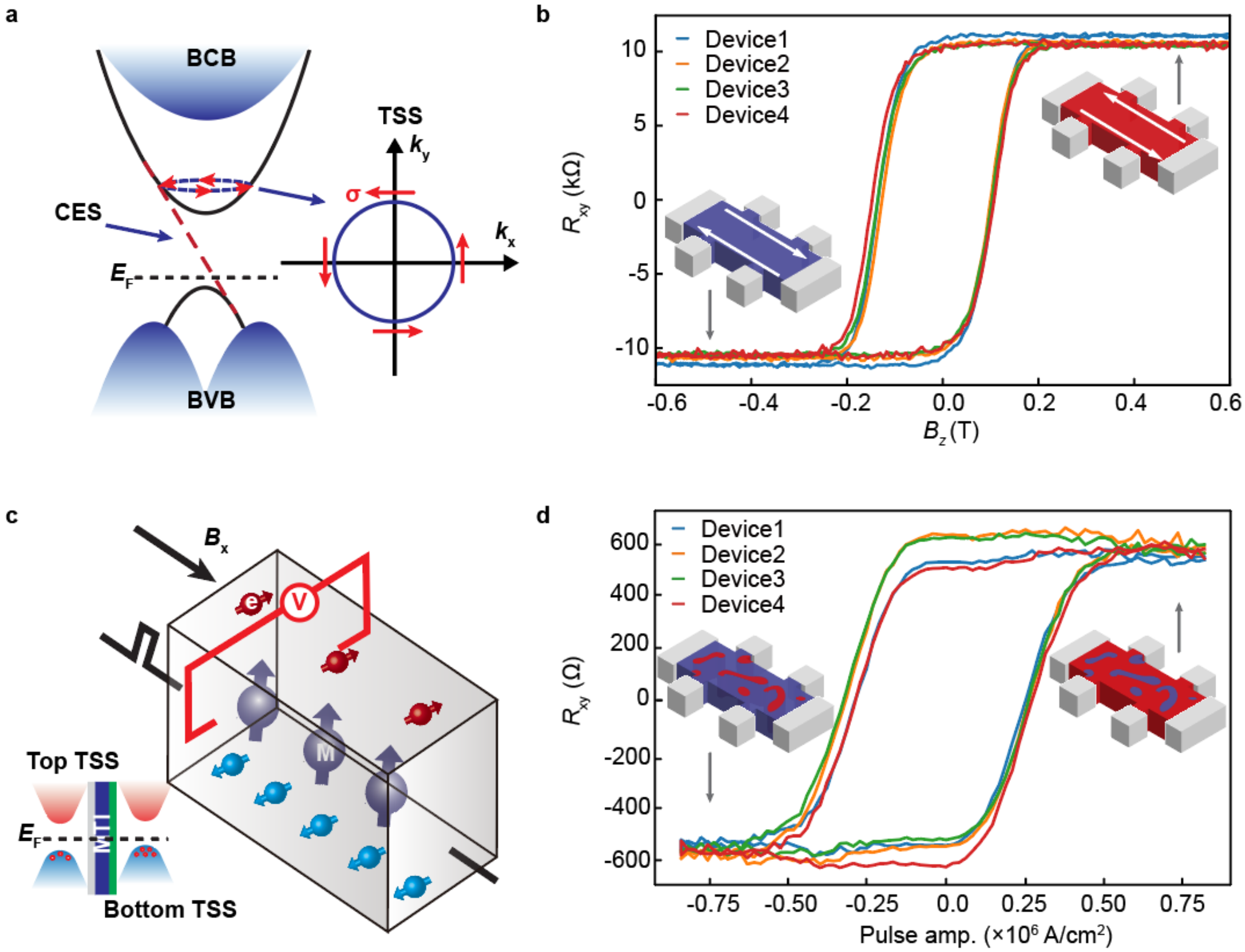

**Figure 1**. **Basic properties of the MTI device**. **a**, The sketch of the band structure of the MTI, where BCB, BVB, TSS, and CES represent the bulk conduction band, the bulk valence band, the topological surface state, and the chiral edge state. The red arrows represent the spin direction on the TSS. The inset shows the spin-momentum locking mechanism of the TSS. Fermi level is indicated (see Supplementary Figure 1). **b**, The out-of-plane magnetic field switching of the four MTI devices D1-D4 used in this work. The left and right insets illustrate the expected magnetic domain state and chiral edge conduction when the magnetization is fully switched down and up, respectively. **c**, The mechanism of the current induced SOT switching of MTI. The red and light blue arrows indicate the spin accumulation generated by the top and bottom TSS, and the blue arrow indicates the local magnetic moment. Left bottom inset shows the schematics of surface bands at the top interface ($AlO_x$/MTI, grey) and the bottom interface (MTI/GaAs, green), resulting in an asymmetric carrier distribution of excited holes. **d**, The pulse write current-induced switching of the four MTI devices. The left and right insets illustrate the expected magnetic multi-domain states when the MTI is switched down and up, respectively. **b** and **d** are obtained using pulse measurements and the amplitude of the reading current is 83 and $8.3\times10^3$ A/cm$^2$, respectively. There is a 30 mT magnetic field along the x-direction for all current-induced SOT experiments.

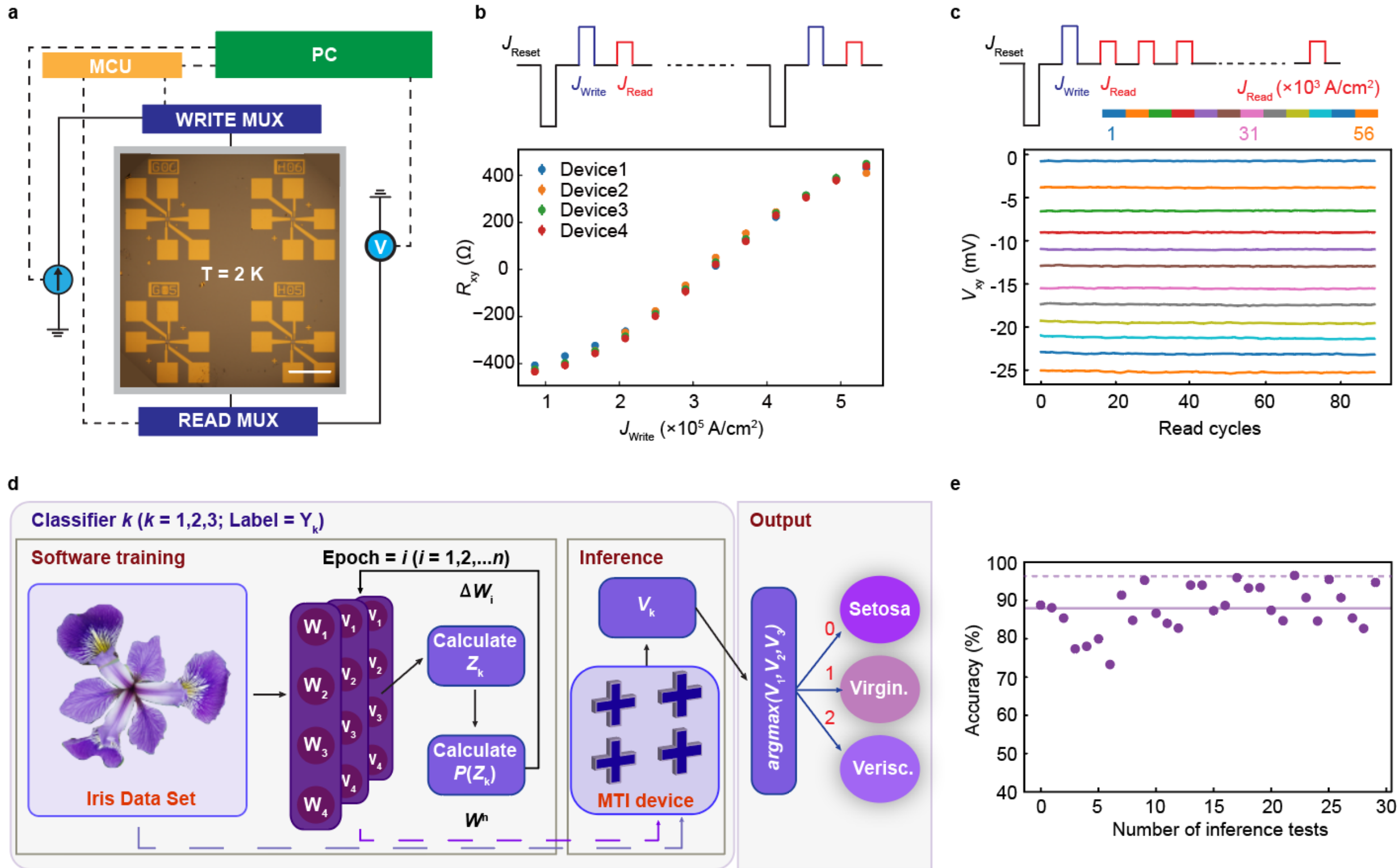

**Figure 2**. **Memristive behavior of the MTI and Iris flower classification. a**, The experimental setup for conducting the inference test with four MTIs. PC, MCU, and MUX are short for personal computers, microcontrollers, and multiplexers/demultiplexers, respectively. The inset in the middle is the optical image of the four MTIs with a channel width of 20 μm, and length of 40 μm. The scale bar is 300 μm. **b**, Lower: the writing curve of 4 MTIs D1-D4. Upper: the scheme of the writing test, where a reset pulse (-8.3 × 10$^5$ A/cm$^2$), a writing pulse, and a reading pulse (2.5 × 10$^4$ A/cm$^2$) are applied in sequence. The Hall resistance is averaged from the reading pulse after 50 trials and the error bar is the standard deviation (see Supplementary Note 4 for details). **c**, Lower: the reading test of the MTI. Upper: the test scheme, where a reset pulse (-8.3 × 10$^5$ A/cm$^2$), a writing pulse, and 90 reading pulses with an amplitude ranging from 8.3 × 10$^2$ A/cm$^2$ to 4.7 × 10$^4$ A/cm$^2$ are applied in sequence. **d**, The diagram of the multi-class cross-entropy algorithm and device inference. **e**, Classification accuracies of 30 inference tests. The dashed and solid horizontal lines indicate maximum and average accuracies of 96% and 88%, respectively.

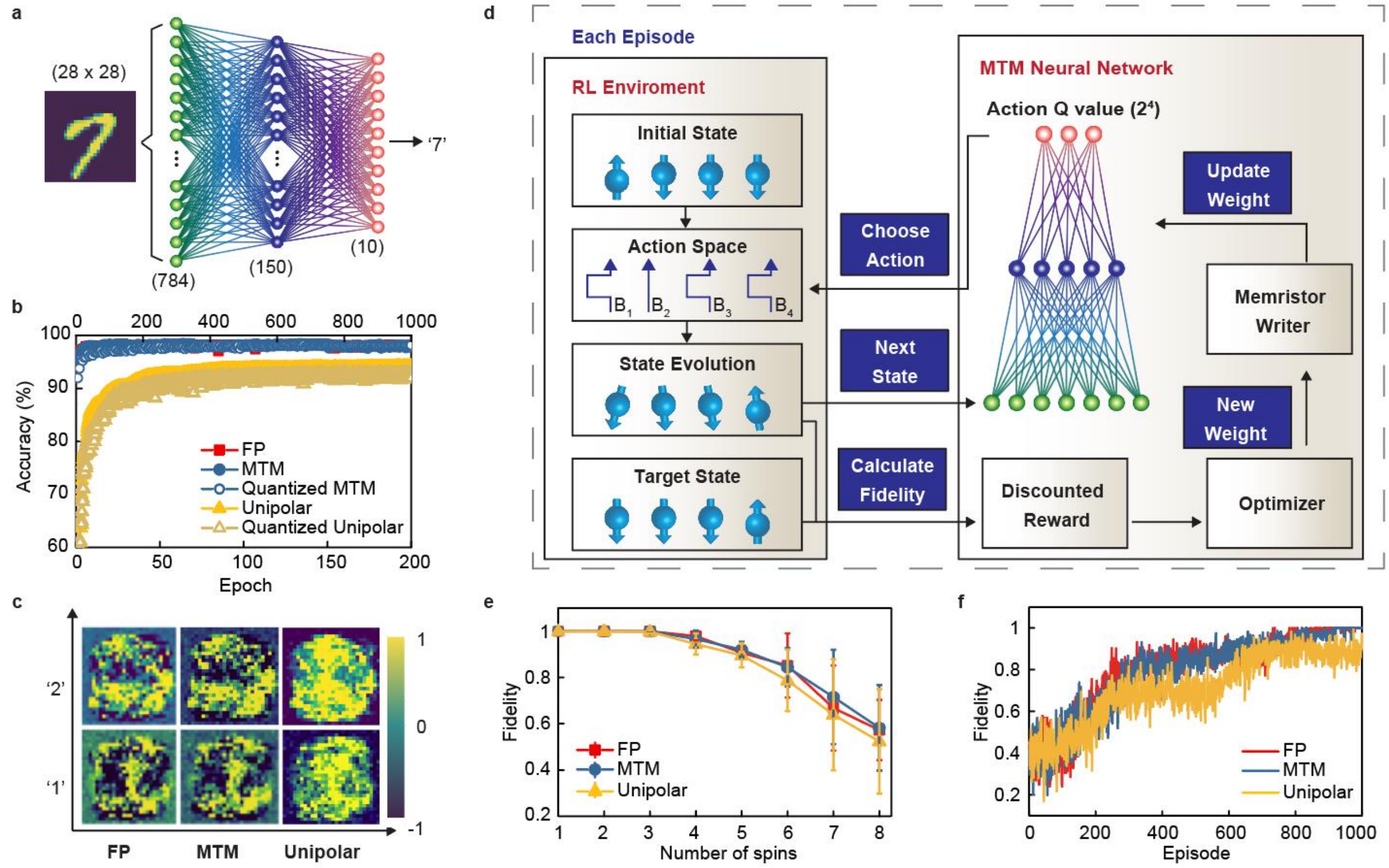

**Figure 3**. **Image recognition and quantum state preparation with MTM neural network**. **a**, The structure of the neural network for the MNIST image recognition. **b**, The image recognition accuracy of floating-point, MTM, unipolar, and quantized neural networks (See Supplementary Table 3 for details of training parameters and schemes). **c**, The weight distribution of three kinds of neural networks after training. **d**, The diagram of the policy gradient algorithm for qubit quantum state preparation (see Methods 'Qubit preparation with policy gradient' for details). **e,** The dependence of average fidelity on the number of spins for different types of neural networks. The fidelity is averaged over at least 12 trials and the error bar represents the standard deviation. **f,** The dependence of average fidelity on the training epoch for different types of neural networks in the case of 3 spins.

## Methods

### Device fabrication and characterization

High-quality single-crystalline Cr-BST films with a thickness of 6 nm were grown on semi-insulating ($\rho \geqslant 10^6\ \Omega\cdot$cm) GaAs (111)B substrates in an ultra-high vacuum Perkin-Elmer MBE system. Before the growth, the substrates were annealed to 580 °C to remove the native oxide, under Te rich environment. High-purity Bi (99.9999%), Te (99.9999%), Cr (99.99%) and Sb (99.999%) were evaporated by conventional effusion cells and cracker cells. During the growth, the substrate was maintained at 200 °C, while the Bi, Sb, Te and Cr cells were kept at 472 °C, 372 °C, 340 °C and 1090 °C, respectively. The cell temperatures of Bi, Sb, and Cr, thus their flux ratio, were fine-tuned to achieve a composition ratio of Cr:Bi:Sb = 0.15:0.26:0.59 with the desirable Fermi level and surface magnetic exchange gap. The epitaxial growth was monitored by an in-situ reflection high-energy electron diffraction (RHEED) technique. The atomically flat surface was evidenced by streaky RHEED patterns, and a growth rate of 1 quintuple layer (QL) per 60 sec (i.e., the lattice spacing of each QL ~1 nm) was measured by the RHEED intensity oscillation in time domain. After the film growth, a 2 nm Al was evaporated to passivate the surface at room temperature.

The Hall bar device is fabricated by the following steps: 1. photolithography to define the Hall bar pattern; 2. ion-beam etching to form the Hall bar structure; 3. photolithography to define the Ti/Au electrode pattern; 4. evaporation and lift-off to form gold electrodes. **Fig. 2a** illustrates the schematic of the device. The MTI Hall bar device has a channel width of 20 μm, length of 40 μm, and thickness of 6 nm. The distance between the centers of two Hall probes is 13 μm. Unless specified, all MTI devices used in this study have the abovementioned dimensions.

The sample is placed in a cryogenic system from Cryogenic Ltd for low temperature measurements; The current pulse is applied by a Keithley 6221 current source; The Hall voltage is measured by a Keithley 2182a nanovolt meter; The controlling program is written by Python, and PyVisa package is used for device communication. All experiments in this paper are conducted at 2 K unless specified. We first characterize the device by applying sequential writing and reading pulses of 2 ms in the x-direction and collecting the Hall voltage in the y-direction. The gap time between writing and reading pulses is 100 ms, which is sufficient for heat dissipation induced by the writing pulse. During the writing phase, a 30 mT symmetry-breaking magnetic field is applied along the x-direction.

The temperature dependence of out-of-plane $R_{xy}$ hysteresis loops is shown in **Extended Data Fig. 1a**. The dependence of $R_{xx}$ is shown in **Extended Data Fig. 1b**. The data is acquired by a current density of $8.3 \times 10^3$ A/cm$^2$. The Arrott plot is used to extract the Curie temperature as shown in **Extended Data Fig. 1c**, where a $T_c$ of at least 35 K can be obtained.

To examine the current amplitude-dependent anomalous Hall resistance (AHR) and Joule heating effect, we carry out current amplitude dependence measurements of out-of-plane hysteresis $R_{xy}$ and $R_{xx}$ loops, which are shown in **Extended Data Figs. 2a and 2b**. By comparing the temperature and current dependence of saturated AHR in **Extended Data Fig. 1d**, we estimate the Joule heating effect ($4.2 \times 10^5$ A/cm$^2$) temporarily heats the sample to ~20 K.

### Co-existence of CES and TSS in MTI

In the ultralow temperature (≤ 260 mK) where the surface states and bulk states freeze out, our Cr-BST exhibits quantized AHR, $h/e^2$ ~25.8 kΩ, due to the transport of pure chiral edge states[33,38]. Meanwhile, at the slightly elevated temperature, while the chiral edge states (CES) still exist, the AHR is not quantized due to the electron conduction from thermally excited topological surface states (TSS) and bulk states (**Fig. 1a**). At 2 K as shown in **Extended Data Fig. 2**, the tangent of (anomalous) Hall angle, which is defined as $R_{xy}/R_{xx,\square}$ (AHR/sheet resistance), achieves a value of 0.5 with a current density of $8.3 \times 10^3$ A/cm$^2$, and increases to 0.6 with a reduced current of $8.3 \times 10^2$ A/cm$^2$ for device D1. This large tangent of Hall angle suggests a non-negligible CES contribution in the MTI memristor as such a high value of the tangent of Hall angle is never reported in non-topological or non-CES-based materials[42,43]. A tangent of Hall angle of 0.6 corresponds to an anomalous Hall angle of 31°. This is around 1/3 of the anomalous Hall angle of 90° that corresponds to the quantum anomalous Hall effect (QAHE) case, where the tangent of Hall angle is infinite. When we further reduce the applied current to 0.833 A/cm$^2$ (1 nA) in our sample, we can observe a drop of $R_{xx}$ when the temperature decreases from 1.81 K to 1.7 K (see **Extended Data Fig. 3a**). Correspondingly, the largest achieved tangent of Hall angle is 0.87 at 1.7 K for D1 (**Extended Data Fig. 3b**). In addition, we have measured other 20-μm samples D5 and D6 that were grown with the nominally same recipe and D5 achieved QAHE at 100 mK with an applied current of 4.38 A/cm$^2$ (5.25 nA) (**Extended Data Fig. 3c**). This suggests the high quality of our MTI samples. We further analyze the temperature dependence of $\rho_{xx}$ and $\rho_{xy}$ in sample D5, which reveals a turning point in $\rho_{xx}$ (**Extended Data Fig. 3d**). To understand the origin, we convert resistance into conductance tensor and obtain $\sigma_{xx}$ by using $\sigma_{xx} = \frac{\rho_{xx}}{\rho_{xx}^2+\rho_{xy}^2}$ (**Extended Data Fig. 3e**). We can see an insulating behavior in the $\sigma_{xx}$ for the whole temperature range. The band structure of an MTI is shown in the inset of **Extended Data Fig. 3e**, where the Fermi level is inside the magnetic surface gap. When the temperature is below the Curie temperature, there is a magnetic gap in the TSS as revealed in spectroscopy [44] and transport measurements [29,33,38,45], where refs. [33,38,45] are our previous works. Due to the nontrivial topology, there is one CES mode. At the QAHE temperature, there is only one dissipationless CES mode, resulting in zero $\rho_{xx}$ and $\sigma_{xx}$, and quantized $\rho_{xy}$ and $\sigma_{xy}$. At temperatures above the QAHE regime, there are finite number of TSS channels due to thermal excitation. Due to the scattering between two edges in the presence of TSS, the CES becomes dissipative and thus has finite resistance.

We can quantitatively evaluate the CES contribution using a phenomenological circuit model[29]. As shown in **Extended Data Fig. 4a**, we have modeled the TSS by the number (n) of effective channels through TSS, $R_{TSS} = \frac{h}{ne^2}$, where $h$ is the Planck constant and $e$ is the electron charge. CES is dissipationless in the QAHE regime and at elevated temperatures becomes dissipative because of the effective scattering between two edges due to presence of TSS. We model the CES by two resistors: a longitudinal resistor with $R_L = nr\frac{h}{e^2}$, where r is the scattering rate, and a transverse resistor with $R_T = \frac{h}{e^2}$. Then we calculate the $R_{xx}$= $V_{xx}/I$=$\frac{nr}{1+n+n^2r}\frac{h}{e^2}$ and $R_{xy}$= $V_{xy}/I$=$\frac{1}{1+n+n^2r}\frac{h}{e^2}$, where $I$ is the total current. In the QAHE regime, $n$=0 and thus $R_{xx}$=0, $R_{xy}$= $\frac{h}{e^2}$. In the high temperature limit, $n$ is very large and thus $R_L >> R_T$, resulting in nearly zero $R_{xy}$. By solving the measured $R_{xx}$ and $R_{xy}$, we can get n and r. Then, we can get the ratios of $I_{CES}/I$=$\frac{1}{1+n+n^2r}$ and $I_{TSS}/I$=$\frac{n+n^2r}{1+n+n^2r}$. We can see that the CES contribution is 0.36 at our device working temperature 2 K when the reading current is 10 μA (**Extended Data Fig. 4b**). The CES contribution remains finite when the temperature is at 20 K. When the applied current decreases to 1 nA, CES contribution (0.51) is larger than that of TSS at 1.7 K, which is consistent with the $R_{xx}$ decreasing trend in **Extended Data Fig. 3a**. In devices D5 and D6, we see much larger contributions from the CES at 2 K (**Extended Data Figs. 4c and 4d**), which are consistent with their larger tangents of Hall angle at 2 K (**Extended Data Figs. 3d and 3f**).

**Scalability and multi-states of MTI memristors**

Multiple resistance states are essential for memristor functions. Usually, when a magnet is scaled to a nanoscale size that is smaller than the domain size, there will be only two states available due to the single domain nature. To investigate the scalability of MTI, we fabricate MTI Hall cross devices with central dimensions of 500 nm (width) by 500 nm (length) and 120 nm (width) by 200 nm (length). We refer them to 500 nm and 120 nm devices, respectively. Fabricating sub-micrometer size MTI devices is a nontrivial task as electron beam lithography (EBL) will damage the MTI samples[46]. We have used low accelerating voltage (20 kV) to fabricate these two devices. To minimize the electron beam exposure, the dimension of the 120 nm device is estimated by the exposure test for fabricating dummy devices on silicon (see the inset of **Extended Data Fig. 5c**). The out-of-plane field-induced magnetization switching results are shown in **Extended Data Figs. 5a and c**. High anomalous Hall resistances of 14 kΩ for 500 nm device and 9 kΩ for 120 nm device, respectively, and multiple intermediate states can be observed in both cases. We further investigated the stability of these intermediate states in MTI by a reading test. The magnetization is first saturated by applying a large positive magnetic field, then the magnetic field is set to a fixed value, and we apply multiple read pulses to obtain $R_{xy}$ of the sample. To reduce the electrical reading noise, we get the averaged value for 10 reads as one data point. This process is repeated 40 times at every magnetic field we have measured. The used magnetic fields range from -110 mT to 100 mT for the 500 nm device and -200 mT to 400 mT for the 120 nm device, respectively. It can be seen in **Extended Data Figs. 5b and d** that at least 17 (15) states can be well separated in 500 (120) nm devices. For the 120 nm device, the average $R_{xy}$ variation is 124.5 Ω corresponding to a reading noise of 1.4%. For the 500 nm device, the average $R_{xy}$ variation is 164 Ω corresponding to a reading noise of 1.1%. This result is reasonable as a larger device should accommodate more intermediate states. In addition, we should be able to find more well separated states in the range between 0 Ω and 10 kΩ if we apply smaller magnetic field steps for the 500 nm device. We further calculate the mean size of the magnetic domain in our 120 nm device using $\sqrt{A/n}$, where $A$ is the central area of the Hall cross, and $n$ is the number of states. We obtain a value of 40 nm. This value is compatible with previous nano-SQUID observations that the magnetic domain in MTI can be as small as tens of nanometers[30]. Note that this value is just a conserved estimation as more states can be available if the field tuning protocol is optimized.

We also compare our results with literature. Qiu et al[47] and Zhou et al[48] have reported the quantum anomalous Hall (QAH) effect in sub-um devices. Especially, ref.[48] shows that QAH effect can still be preserved in a 72 nm-width Hall bar device, indicating the decaying length of CES is less than 36 nm. More importantly, this 72 nm-width Hall bar device does not show single-domain switching behavior as intermediate states are observed during out-of-plane magnetic field switching. With this previous work and our data, we conclude that MTI memristors have a high potential to scale down to sub-100 nm while still holding multiple magnetization states and giant anomalous Hall resistance.

**Implementation of Iris pattern classification**

The procedure of the Iris pattern classification is shown in **Fig. 2d**. We build three binary classifiers corresponding to the 3 types of flowers (Setosa, Versicolor, and Virginica) in the Iris dataset. The classifiers are identical, training on the same input data $S$ with a dimension of [4× 150], but with the label $Y$ binarized to each flower (e.g., for the Setosa classifier, Setosa samples are labeled as "1" while other samples are labeled as "0"). Each classifier, therefore, has a [1×4] weight matrix $W$, and the output is generated by multiplying the input data with the weight matrix to obtain the prediction $Z=WS$, then converted to a probability $P(Z)$ via a sigmoid function:

$$\boldsymbol{P(Z) = \frac{1}{1 + e^{-Z}}} \quad (1)$$

We adopt the cross-entropy cost function and gradient descent for the optimizer, which corresponds to the weight update rule given by:

$$\Delta\mathbf{W} = \frac{\boldsymbol{\gamma}(\boldsymbol{P}(\boldsymbol{Z}) - \boldsymbol{Y})\boldsymbol{S}^{\boldsymbol{T}}}{\boldsymbol{N}} \tag{2}$$

where $\gamma$ is the learning rate and $N$ is the number of input samples. At epoch $n + 1$, $W^{n+1} = W^{n}-\Delta W$. The final prediction result (PR) is obtained by comparing $P(Z)$ of each classifier:

$$\boldsymbol{PR} = \mathbf{argmax}(\boldsymbol{P}(\boldsymbol{Z}_{\mathbf{Setosa}}), \boldsymbol{P}(\boldsymbol{Z}_{\mathbf{Versicolor}}), \boldsymbol{P}(\boldsymbol{Z}_{\mathbf{Virginica}})). \tag{3}$$

PR can take values of 0, 1, and 2 which correspond to Setosa, Virginica, and Versicolor, respectively. The PR is then compared with the original label of the sample to obtain accuracy.

We then demonstrate the classification of the Iris dataset using the MTI devices. The Iris dataset contains 150 samples, each with the measurements of 4 features of the flower: sepal length, sepal width, petal length, and petal width. We train the network depicted in **Fig. 2d** on the entire Iris dataset using the above logistic regression algorithm. In our training, we preprocess the original data to connect the software training with our memristor network. We first translate the iris input data to a language that the device can recognize (e.g the current): scale the iris input to a range from 2 to 4 through a normalization function, which corresponds to a range from 20 μA to 40 μA for the input of the memristor. The final accuracy is 96%, as shown in **Extended Data Fig. 6a**. The weight matrices for each flower are:

Classifier 1: $W_{setosa} = [2.03, 10.02, -11.24, -1.93]$,

Classifier 2: $W_{versicolor} = [2.02, -2.94, 0.63, -0.68]$,

Classifier 3: $W_{virginica} = [-6.63, -12, 13.75, 4.93]$.

We then map the weights to the AHR of the MTM such that its AHR is within the range between -200 Ω and 200 Ω (the writing current has a linear relationship with the AHR), and the input currents to the AHRs are in the range of from 20 μA to 40 μA (the reading current in this range will only bring 2% noise) is:

Classifier 1: $W'_{setosa} = [30.45\Omega, 150.3\Omega, -168.6\Omega, -28.95\Omega]$,

Classifier 2: $W'_{versicolor} = [30.3\Omega, -44.1\Omega, 9.45\Omega, -10.2\Omega]$,

Classifier 3: $W'_{virginica} = [-99.45\Omega, -180\Omega, 206.25\Omega, 73.95\Omega]$.

The inference is conducted in the following steps. Step 1: the AHR of 4 memristors is programmed to $W'_{Setosa}$ to hardware implement classifier 1. Step 2: $I_m$ is sent to the memristor array row by row and the Hall voltages of 4 memristors are measured and summed together. The total Hall voltage of the 4 memristors is the output of classifier 1 (denoted as $V_1$). Step 3: Steps 1-2 are repeated for Classifiers 2 and 3 ($V_2$ and $V_3$). Finally, we apply an argmax function on $V_1$, $V_2$, and $V_3$ to obtain the final classification result (FCR). We compare this with the ideal classification accuracy as well as the simulated accuracy in **Extended Data Fig. 6b**. Simulation across 100 trials results show that the AHR should achieve an average of 88.6% accuracy. Experimentally, the network achieves an average accuracy of 87.8% across 30 inference tests (**Fig. 2e**).

## Qubit preparation with policy gradient

We adopt the quantum-state preparation task described in Zhang et al.[10] (**Fig. 3d**). The task aims to control the state of *K* serially coupled spins from an initial state to the target state via a magnetic field *B*. The initial state is the leftmost spin in the $|1\rangle$ state and all others in the $|0\rangle$ state. The target state has the rightmost spin in the $|1\rangle$ state and all others in the $|0\rangle$ state. A policy gradient system is adopted for the task. The system composes an environment with state *S*, an agent that produces an action *A*, and the reward of the action *R*.

The environment represents the evolution of the *K* serially coupled spins. *S* represents the state of the spins, and is a complex vector of length *K*. The evolution of *S* is determined by the Hamiltonian *H*, which can be computed as

$$\boldsymbol{H(t) = C\sum_{k=1}^{K-1}\left(S_x^k S_x^{k+1} + S_y^k S_y^{k+1}\right) + \sum_{k=1}^{K} 2B_k(t) S_z^k} \tag{4}$$

Where *C*=2 is the coupling strength between adjacent spins and $B_k(t)$ is the control magnetic field at timestep *t*. The evolution of the state follows the Hamiltonian mechanics, e.g.,

$$S(t+dt) = S(t)e^{-iH(t)dt} \tag{5}$$

Where d*t* is the length of the time step.

The agent is a network that produces the control magnetic field *B*. It is a two-layer fully-connected neural network, with an input feature of size *2K* (e.g. the real and imaginary components of the *K* spins), a number of neurons of *N*=16, and an output of size $2^K$. (e.g., all possible actions; $B_{\mathrm{ctrl}} = \pm 40$ for each of the *K* spins; for example, for *K*=2, the possible configurations are [-$B_{\mathrm{ctrl}}$, -$B_{\mathrm{ctrl}}$], [-$B_{\mathrm{ctrl}}$, +$B_{\mathrm{ctrl}}$], [+$B_{\mathrm{ctrl}}$, -$B_{\mathrm{ctrl}}$], and [+$B_{\mathrm{ctrl}}$, +$B_{\mathrm{ctrl}}$]). The agent selects an action with a probability based on the softmax of the neural network output.

The reward is computed based upon a fidelity function *f*, which is the absolute distance between the environment state *S* and the target state $S_t$, e.g.,

$$f = |<S|S_t>|^2 \tag{6}$$

The reward *R(t)* at timestep *t* is equal to $R_{\max}$ =2500 when $f$ is within $\varepsilon$ ($\varepsilon = 0.01$) of its maximum value of 1, indicating that the current state is already in the target state, and *10*f* otherwise.

In each trial, the network is trained for $N_{\mathrm{ep}}$=1000 episodes. During each episode, the environment and agent interact for $N_s$=20 steps. In each step, the agent generates the control magnetic field, the state of the environment evolves, and a reward is computed. At the end of each episode, the episode reward $R_{\mathrm{ep}}$ is obtained via a discounted reward function, e.g.,

$$R_{\mathrm{ep}} = \sum_{n}^{N_S} r^t R(t) \tag{7}$$

where $r = 0.99$ is the discount rate. This reward is the loss used to update the neural network in each episode. When the trial is complete, the network is determined to be trained. We report the maximum $f_{\mathrm{ep}}$ among the last 10 episodes as the fidelity of the trial $f_{\mathrm{trial}}$.

For the floating-point network, we directly train upon a software-based neural network with weights that can take on any value. For our bipolar MTM network, the trained weights in each episode are converted to the resistances of the MTM via the following steps: (a) clipping the weights to the range of [-1, 1], (b) mapping it to the resistance range of [–600, 600], (c) finding the corresponding currents, and (d) applying them to an array of device models with a write error of 2%. The unipolar network follows the same process with the weights clipped to the range of [0,1] and mapped to the resistance range [1000, 3000] instead.

We run the three networks for different configurations involving the number of spins $K$ to control. For each configuration, we execute at least 12 trials and report the average $f_{\mathrm{trial}}$ in **Fig. 3e** and the average fidelity of each episode in **Fig. 3f**.

**Design challenges for Hall effect-based neural network**

Constructing a practical AHR neural network confronts huge challenges in both materials and network designs. From the material perspective, the AHR of normal heavy metal (HM) and ferromagnetic transition metal (FM) structures is less than 5 Ω[49,50]. For an HM/FM Hall bar device with a channel width of 100 nm and a thickness of 5 nm, the required current density for generating reliable readout signals (e.g., to surpass the thermal and coupling noise of transistors, which is in the range of millivolts) is more than $10^8$ A/cm$^2$. This current is an order of magnitude higher than the switching current and leads to disturbance in the device state. Meanwhile, the SOT efficiency in these structures also needs improvement, being capped at 1. In this context, MTI possesses incomparable advantages owing to its 2-3 orders higher AHR and SOT efficiency. MTM-based neural network is thus expected to exhibit much lower power consumption.

We make a note of our noise consideration in circuit design. For fast and reliable NN operation, the circuit requires that the minimal current step is on the order of a few µA and voltage step on the order of several mV. This is because even if variations are compensated, there are still components difficult to account for: (a) leakage current of unselected paths during read operation (usually on the order of few hundred nA), and (b) thermal noise (on the order of a few hundred µV @ 100MHz).

From the network perspective, a direct connection of any of the four terminals of a Hall bar could lead to leakage/sneak current paths that not only cause large energy consumption but degrade the readout signal. Since the Hall bar device is a 4-terminal device in which every terminal is conductive, there are many paths where leakage current can flow. However, this issue has yet to be considered and investigated. Recently, Lan et al. have proposed to connect the Hall bar devices in series to sum Hall voltages[51]. Below, we show that this design will be problematic and fail to sum AHE voltages correctly using COMSOL simulation.

The AHE voltage of a single Hall bar device is shown in **Extended Data Figs. 8a** and **b**. The conductivity of the magnetic material and the connecting wires are $\sigma_{\mathrm{mag}} = \begin{pmatrix} 10^4 & 10^3 \\ -10^3 & 10^4 \end{pmatrix}$ S/m and $\sigma_{\mathrm{metal}} = \begin{pmatrix} 10^9 & 0 \\ 0 & 10^9 \end{pmatrix}$ S/m, respectively. The color on the contour plot represents the electric potential of the device when a 10 V voltage is applied. **Extended Data Fig. 8b** shows the dependence of Anomalous Hall voltage (AHV) on the applied voltage. As expected, the AHV is proportional to the AHR and the input voltage.

We then simulate 3 Hall bar devices with their Hall channels in series as shown in **Extended Data Fig. 8c**. The devices have different AHRs translating to different diagonal conductivities in the simulation, as,

$$\sigma_{\mathrm{mag1}} = \begin{pmatrix} 10^4 & 2 \times 10^3 \\ -2 \times 10^3 & 10^4 \end{pmatrix} \text{ S/m;}$$

$$\sigma_{\mathrm{mag2}} = \begin{pmatrix} 10^4 & 3 \times 10^3 \\ -3 \times 10^3 & 10^4 \end{pmatrix} \text{ S/m;}$$

$$\sigma_{\mathrm{mag3}} = \begin{pmatrix} 10^4 & -3 \times 10^3 \\ 3 \times 10^3 & 10^4 \end{pmatrix} \text{ S/m.}$$

The input voltages are U, 2U, and 3U, respectively. We compare the AHV of each of the devices individually, then compare their sum with the voltage of 3 Hall bars in series (**Extended Data Fig. 8d**). It is clear that $V_t \neq V_{xy1}+V_{xy2}+V_{xy3}$.

This simulation result suggests that a simple connection of Hall bar devices fails to sum Hall voltages. Now we discuss why it happens. The electric potential contour plot is shown in **Extended Data Fig. 8c**. Considering only the vertical biases, the potential in the center of each Hall bar should be 0.5U, U, and 1.5U. Considering only the lateral biases, the center potential of each Hall bar should be the center of the adjacent Hall bar plus half the AHV of the two to enable summation. The disagreement in the potential difference in the vertical and horizontal paths can be viewed as a leakage path between the two paths that impact the summation of the AHVs, therefore, rendering the summation problematic.

As shown in **Extended Data Fig. 8e**, we also consider applying antisymmetric voltages across the Hall bar device so that the center potential between each device is smaller. The input voltages for each Hall bar device are 0.5U and -0.5U, U and -U, 1.5U and -1.5U, respectively. The conductivity of the magnetic material and the wires is the same as in the previous simulation. As shown in **Extended Data Fig. 8f**, the AHV of 3 Hall bars in series is still not equal to the summation of the individual AHV of each Hall bar, although the difference is much smaller than in the previous case. It is worth mentioning that this leakage path also impacts the write operation as the designated current density passing through the channel is changed. Therefore, making sure that each device operates the same when they are independent and when they are in series is crucial to correct neural network operation.

**Verification of Hall current mode reading**

Experimental characterization of AHC has not been reported despite extensive research on AHV. Hence, we start by characterizing AHC in a Hall bar device where voltages are applied to both longitudinal and transverse channels. We first study AHC via a Finite Element method. **Extended Data Fig. 9a** shows the device model of the simulation where the color represents the voltage potential. The voltages are applied to the two ends of the longitudinal (x-) channel with the same amplitude but opposite signs so that the center of the device is a virtual ground. The simulation result suggests the y-channel current presents very similar behavior as the AHV. A linear relationship between $I_y$ and magnetization (**m**) is observed in **Extended Data Fig. 9b**, where **m** changes the off-diagonal conductivity. The terminal current passing through the y-channel, $I_y$, is proportional to U as shown in **Extended Data Fig. 9c**.

We then conduct experiments to characterize AHC in an MTI Hall bar device with an AHR of about 8000 Ω. The hysteresis loop of the device is shown in **Extended Data Fig. 9d** with the device photo shown in the inset. The structure of the device and the experiment configuration are shown in **Extended Data Fig. 9e**. The voltages are applied by 2 Keithley 2450 source meters. AHR and AHC are measured by a Keithley 2000 multi-meter. $U_1$ is swept from 10 mV to 100 mV while keeping $U_2$ about 7× of $-U_1$ to keep the potential at the intersection of the x and y channels near zero. It can be seen in **Extended Data Fig. 9f** that there is a linear relationship between $U_1$ and the Hall voltage and $I_y$, and after reversing **m** the slope of the curve reverses sign. This phenomenon agrees with **Extended Data Figs. 9b and c** and the numerical values also fit the model. The resistance of the side channel is 52 kΩ. The main channel current (~12.5 μA at $U_1$ = 100 mV) is about 6 times of AHC while the side Hall bar resistance is also about 6 times of AHR which agrees with the simulation.

We then empirically obtain the following relationship according to the above observations,

$$I_y = \frac{V_x R_{\mathrm{H}}}{R_{sx} R_{sy}} \tag{8}$$

where $R_{sx}$ and $R_{sy}$ are the two-terminal resistance of the x-channel and y-channel of the Hall bar, respectively, and $R_{\mathrm{H}}$ is the Hall resistance of the device. This AHC retains the proportional dependence on $R_{\mathrm{H}}$ and $V_{\mathrm{x}}$.

Having established the characteristics of AHC in a single Hall bar device, we extend our model to parallel-connected Hall bar networks. The equivalent circuit model is shown in **Extended Data Fig. 10a**. Each Hall bar can be modeled as a voltage source with an electromotive force (EMF) of $V_{\mathrm{n}}$ equal to its AHV and transverse channel (y channel) resistance of $R_{sy}$. When the devices are connected in parallel, the total current $I_{\mathrm{t}}$ will be equal to the sum of the terminal current according to Kirchhoff's law, e.g.

$$I_{\mathrm{t}=} I_1 + I_2 + \cdots I_n \tag{9}$$

where $n$ is the device number, $I_i$ ($i$ =1, 2…, $n$) is defined as the AHC of each device. As each Hall bar device and the ampere meter are connected in parallel, the terminal voltage for all Hall bars is zero. We can thus calculate AHV ($V_i$) of each device via Eq. (8) and obtain,

$$V_i = I_i R_{sy} \tag{10}$$

Combining Eq. (5) and (6), we can obtain

$$V_1 + V_2 + \cdots V_n = I_{\mathrm{t}} R_{\mathrm{sy}} \tag{11}$$

We thus obtain,

$$I_{\mathrm{t}} = \frac{V_1 + V_2 + \cdots V_n}{R_{\mathrm{sy}}} \tag{12}$$

This equation suggests the linear summation of AHV can be represented by measuring the total terminal current $I_{\mathrm{t}}$. In other words, the current mode reading scheme can be applied to readout VMM operations in neural networks like AHV. Because each Hall bar is driven independently and the potential across the y channel is the same, Hall signals can be correctly summed without the leakage current issue.

To verify our model, we confirm the summation of AHC using COMSOL simulation. We put 3 Hall bar devices in parallel as shown in **Extended Data Fig. 10b**. The top and bottom bus lines are connected so that the current on the bus line can be measured. The conductivity of connecting wires and each Hall bar is the same as in the section (Methods "Design challenges for Hall effect-based neural network'). The terminal voltages are 2U, 4U, and 6U. The total current on the bus line ($I_{\mathrm{t}}$) is presented in **Extended Data Fig. 10c**. We compare the results with the individual AHC of each device ($I_1$+$I_2$+$I_3$) and confirm that the results match Eq. (12). We also successfully verify the summation of AHC signals in a circuit-level simulation (see next section for detail). These results verify the feasibility of summing AHC signals for VMM in the AHR neural network.

It is worth noting that Yang et al. also presents a design that connects Hall bar devices in parallel for neural network operation[52]. Their design reads the terminal voltage of parallel connected devices for VMM. The terminal voltage $U_{\mathrm{t}}$ will be the average AHV of all devices, e.g., $U_{\mathrm{t}} = \frac{V_1 + V_2 + \cdots V_n}{n}$. The issue in this design is that the contribution of each device voltage is averaged, prohibiting the development of large neural network arrays (the signal of each device is divided by $n$). On the other hand, our design maintains the output signal.

## Data Availability

The data that support the plots within this paper and other findings of this study are available at DataSpace@HKUST Digital Repository [53].

## Code availability

Other than commercial software, the codes used for this study can be found at DataSpace@HKUST Digital Repository [53].

## Methods-only references

**Extended figures**

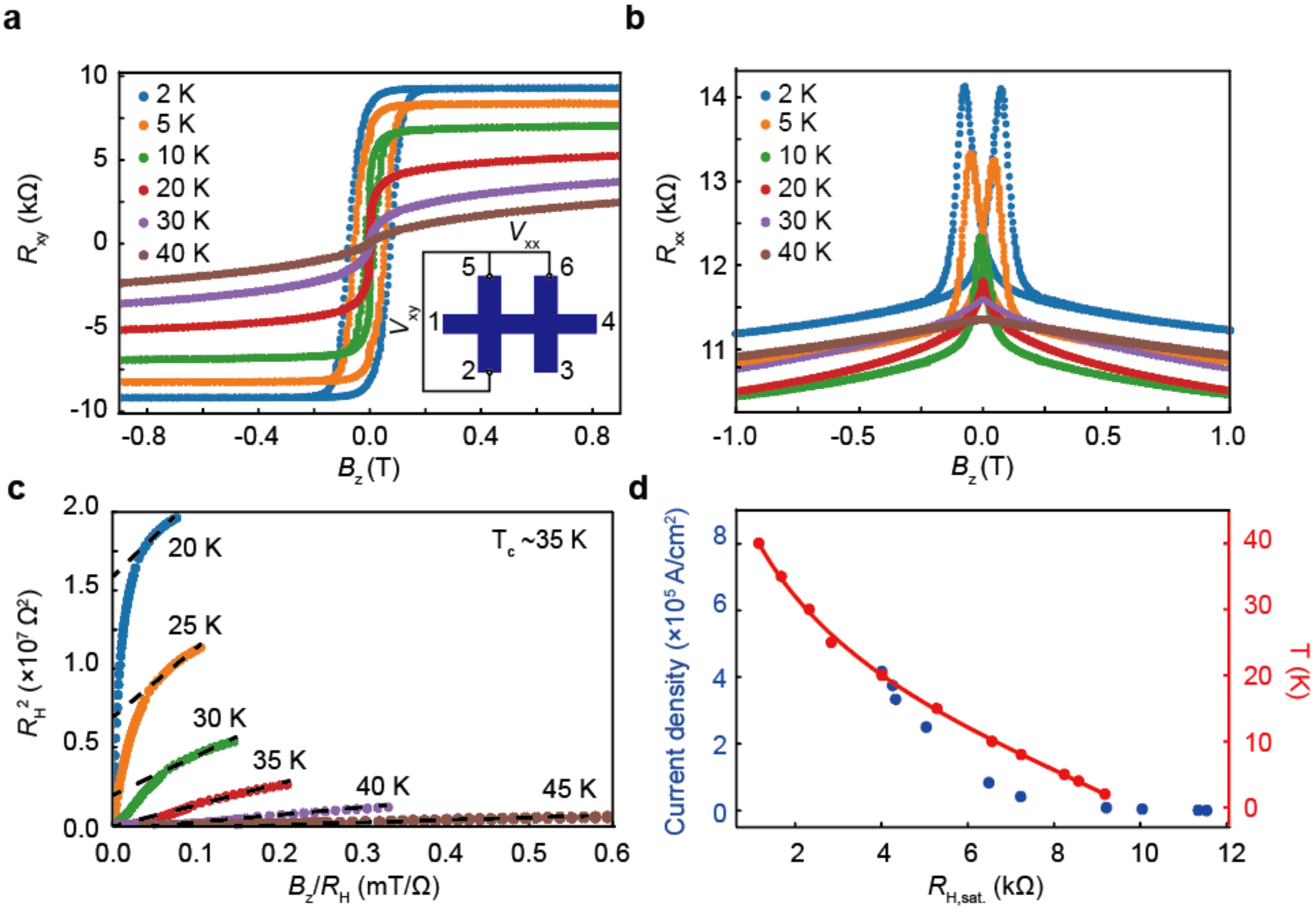

**Extended Data Fig. 1. Temperature dependence of $R_{xy}$ and $R_{xx}$ of the MTI device D1. a** and **b**, $R_{xy}$ and $R_{xx}$ as a function of magnetic field at different temperatures. **c**, Arrott plot for the MTI device. **d**, The relationship between current density and saturated AHR, and the relationship between temperature and saturated AHR. The data in a-c is acquired by a reading current of $8.3 \times 10^3$ A/cm$^2$ (10 μA).

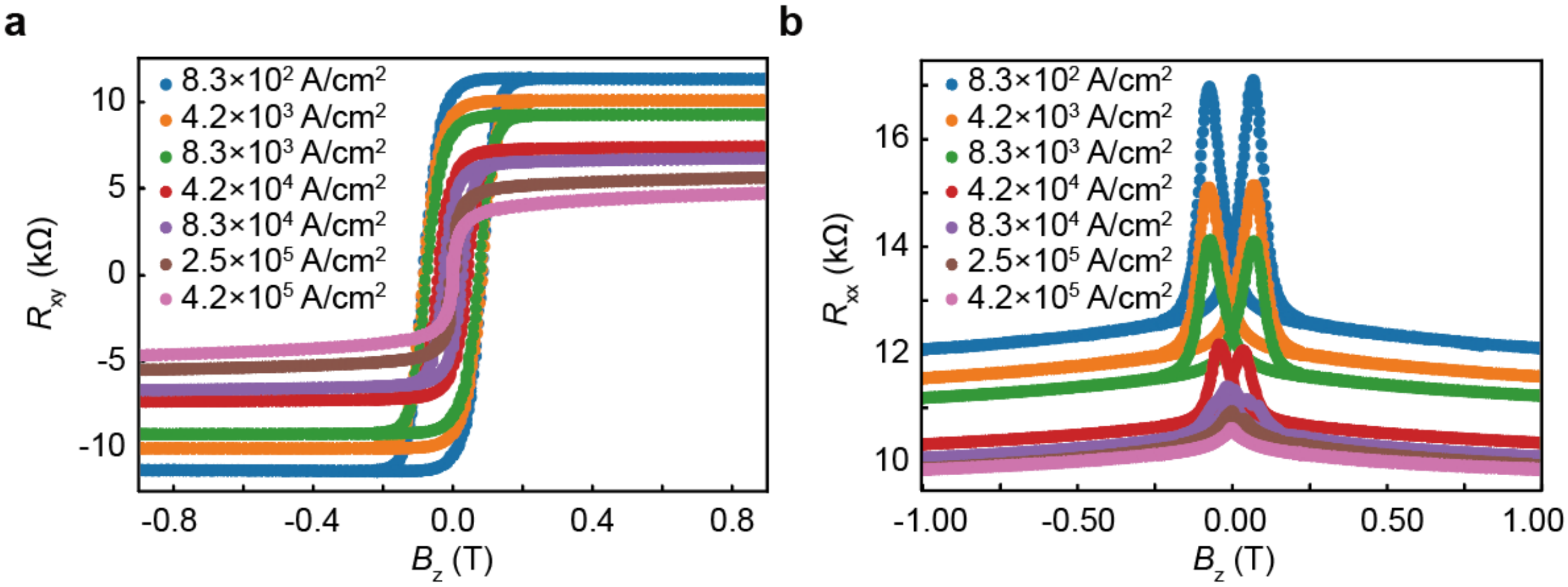

**Extended Data Fig. 2. Current dependence of the MTI device D1.** Current dependence of $R_{xy}$ (**a**) and $R_{xx}$ (**b**) as a function of out-of-plane magnetic field at 2 K for the MTI device D1.

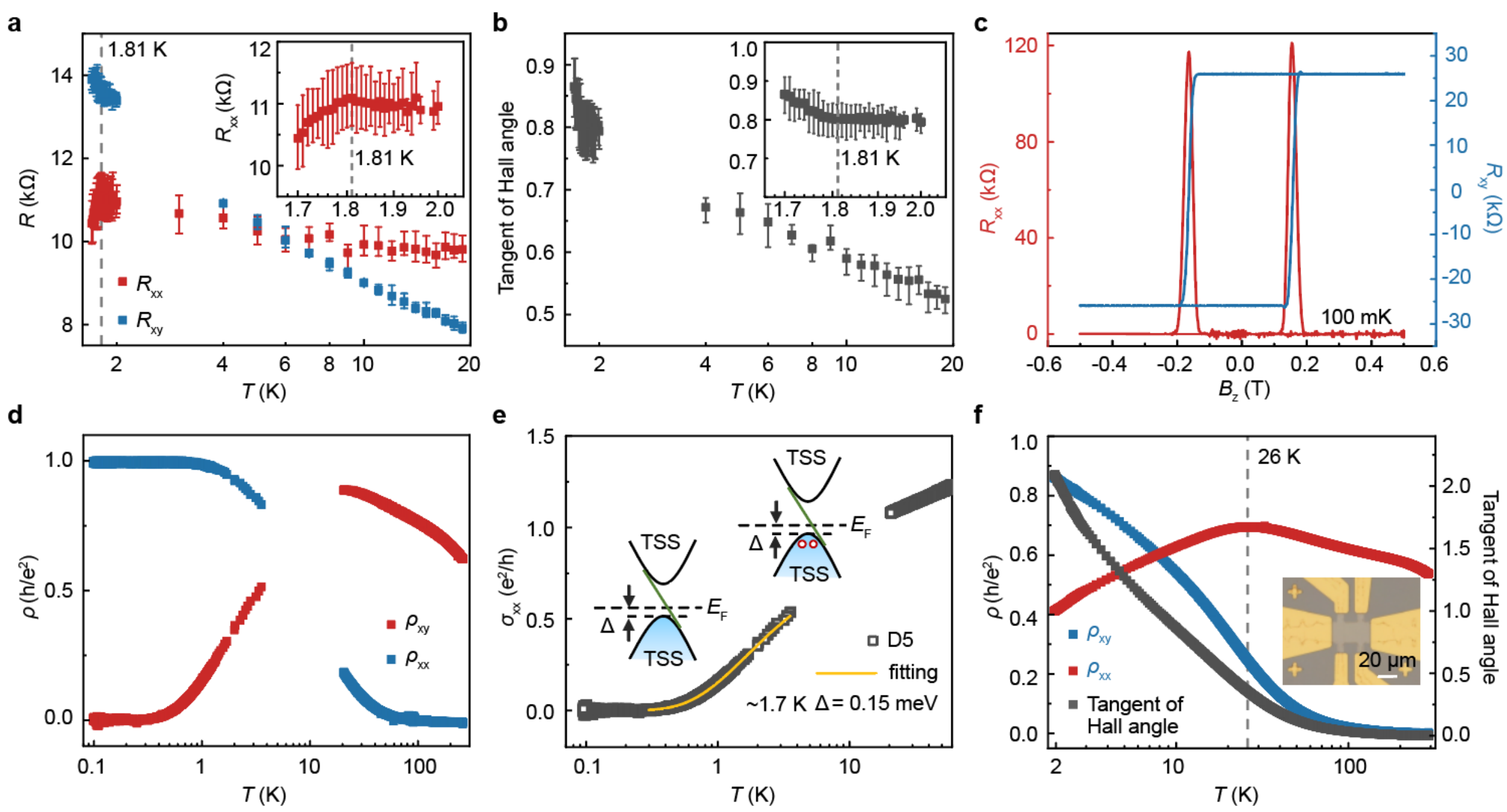

**Extended Data Fig. 3. Analysis of temperature-dependent magneto-transport data in MTIs. a**, Longitudinal resistance $\rho_{xx}$ and transverse resistance $\rho_{xy}$ of an MTI as a function of temperature in D1. The reading current is 0.833 A/cm$^2$ (1 nA). **b**, Tangent of Hall angle in D1. The upper and lower limits of error bars in a and b represent the maximum and minimum values, respectively. **c**, $R_{xx}$ and $R_{xy}$ as a function of magnetic field of another 20-μm sample D5 (with the nominally same growth recipe) showing QAHE at 100 mK. The reading current is 4.38 A/cm$^2$ (5.25 nA). **d**, Longitudinal sheet resistance $\rho_{xx}$ and transverse resistance $\rho_{xy}$ of D5 as a function of temperature. Note that some data points are not presented due to technical errors during the measurements. **e**, Longitudinal sheet conductance $\sigma_{xx}$ of D5 as a function of temperature. Insets show schematics of band structure of TSS and CES in the MTI below QAHE temperature (left) and above QAHE temperature with excited holes from TSS (right). **f**, Longitudinal sheet resistance $\rho_{xx}$, transverse resistance $\rho_{xy}$ and tangent of Hall angle of D6 as a function of temperature measured by 833 A/cm$^2$ (1 μA).

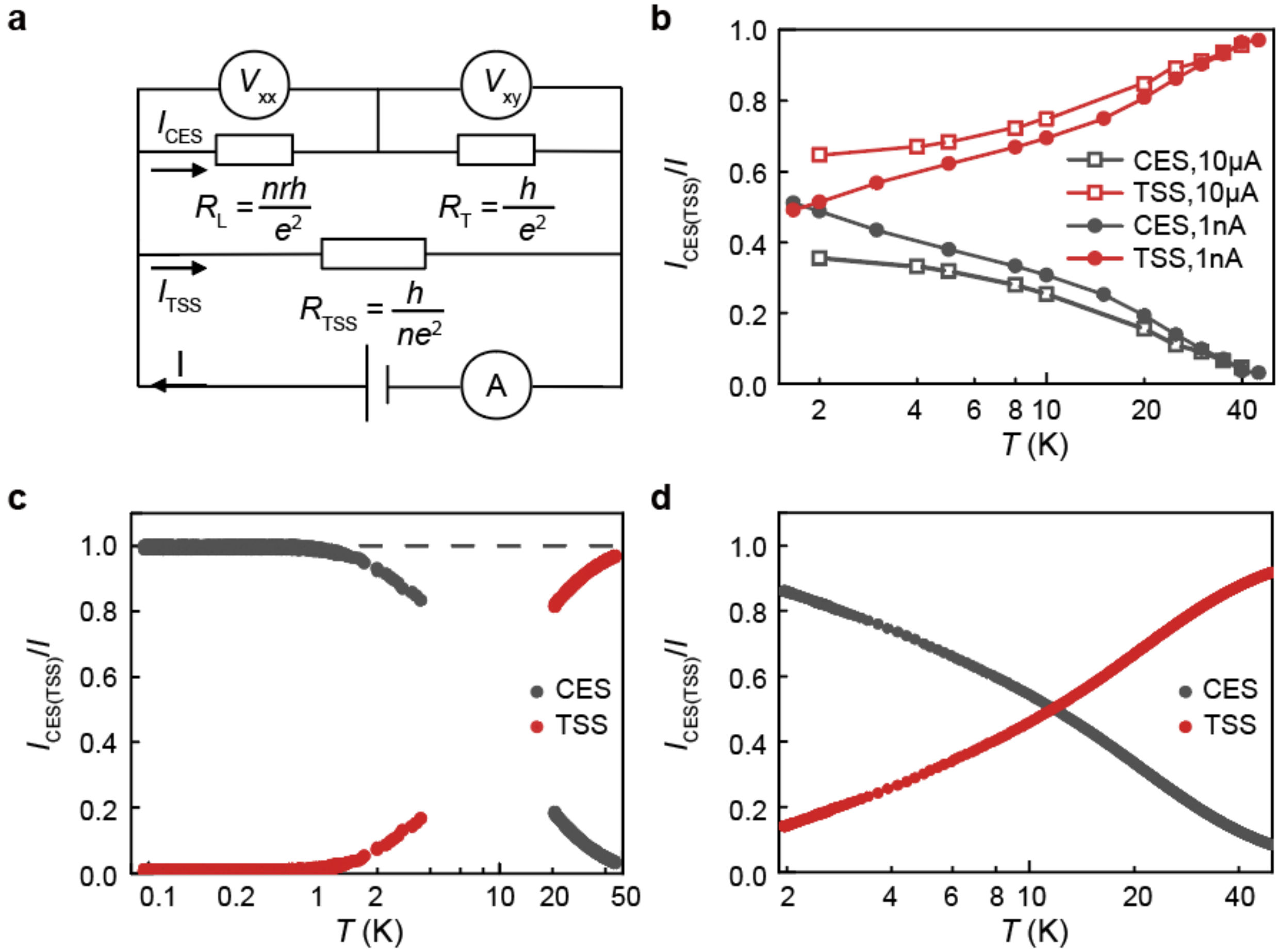

**Extended Data Fig. 4. Quantifying contributions of TSS and CES in MTIs. a**, Phenomenological circuit model for describing the TSS and two CES in MTI samples; $R_L$ describes the longitudinal resistance, $R_T$ describes the anomalous Hall transverse resistance, and $R_{TSS}$ is the TSS contribution. $I_{CES}$ and $I_{TSS}$ are the current going through the CES and TSS, respectively. $N$ is the number of effective channels for TSS, and $r$ is the scattering rate between CES and TSS. **b**, Calculated contributions of CES and TSS as a function of temperature in D1. The reading current is 10 μA and 1 nA, respectively. **c**, Calculated contributions of CES and TSS as a function of temperature in D5. The reading current is 5.25 nA. **d**, Calculated contributions of CES and TSS as a function of temperature in D6. The reading current is 1 μA.

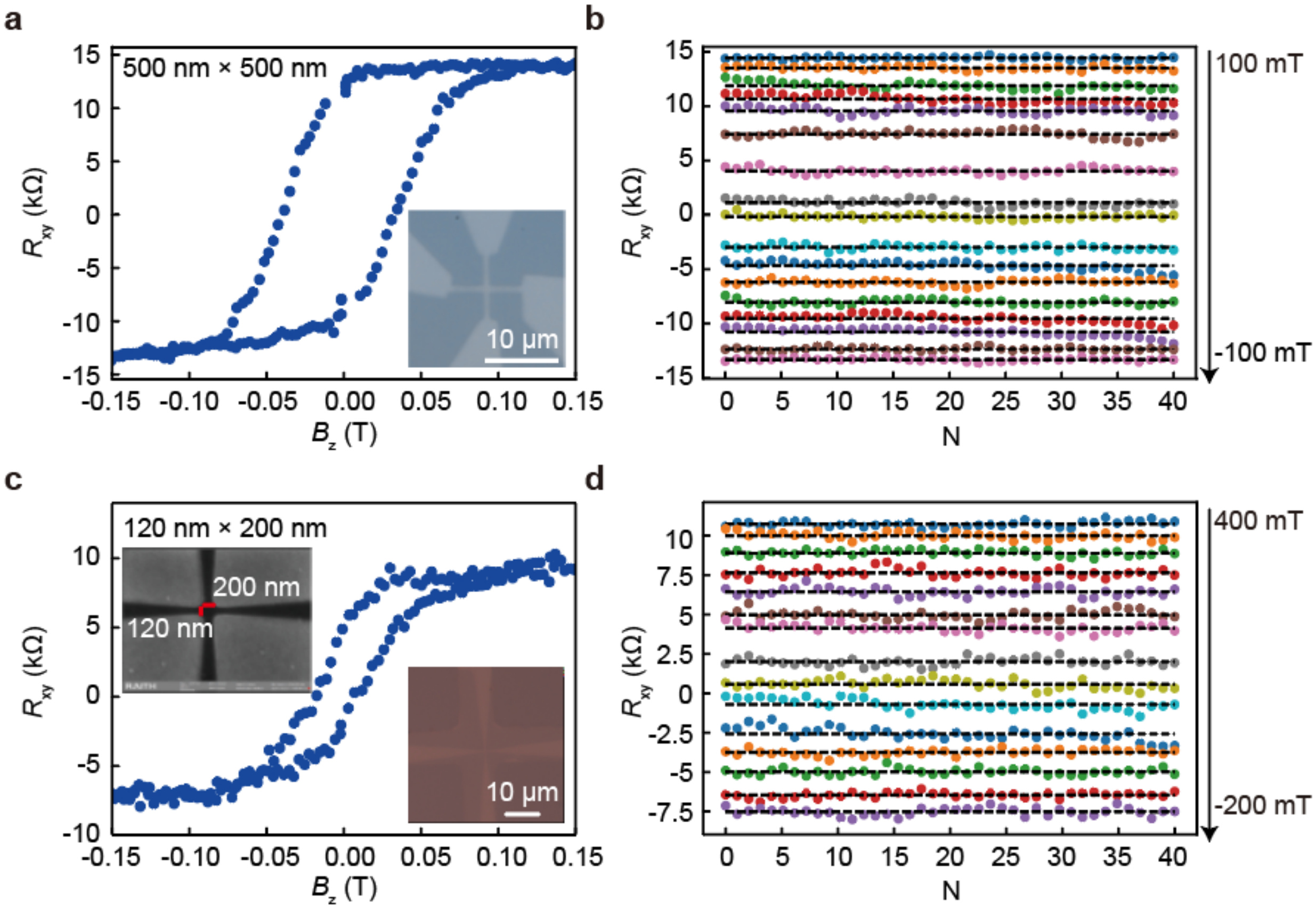

**Extended Data Fig. 5. Multi-states in 500 nm and 120 nm MTI devices**. **a**, Field induced magnetization switching in an MTI device with a dimension of 500 nm by 500 nm. The inset shows the optical image of the MTI device and the reading current is 75 nA. **b**, The reading test of a 500 nm MTI device. $R_{xy}$ is read for 40 times at magnetic fields ranging from −110 mT to 100 mT. **c**, Field induced magnetization switching in an MTI device with a dimension of 120 nm by 200 nm. The reading current is 30 nA. The upper inset shows the SEM image of the exposure test result on Si substrate. The lower inset shows the optical image of the MTI device. **d**, The reading test of a 120 nm MTI device. $R_{xy}$ is read for 40 times at magnetic fields ranging from −200 mT to 400 mT.

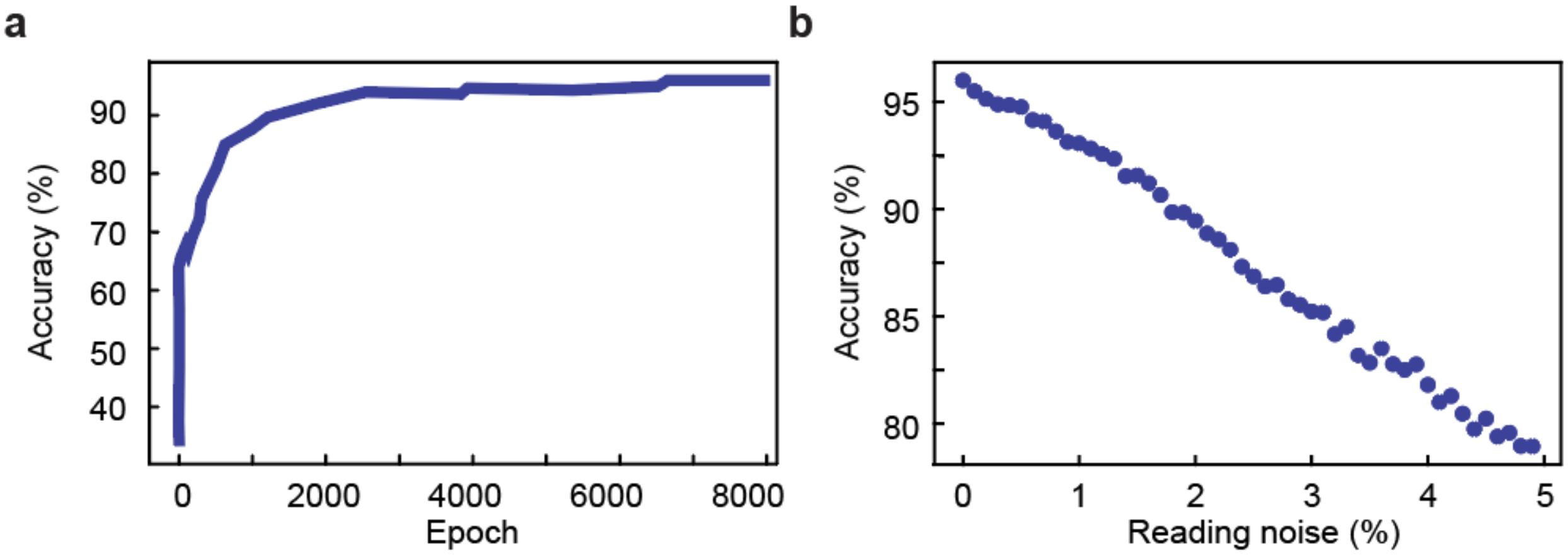

**Extended Data Fig. 6. Simulation result of iris pattern classification. a**, The dependence of classification accuracy on the training epoch. **b**, The dependence of accuracy on the reading noise of MTM. The accuracy is averaged from 100 trials.

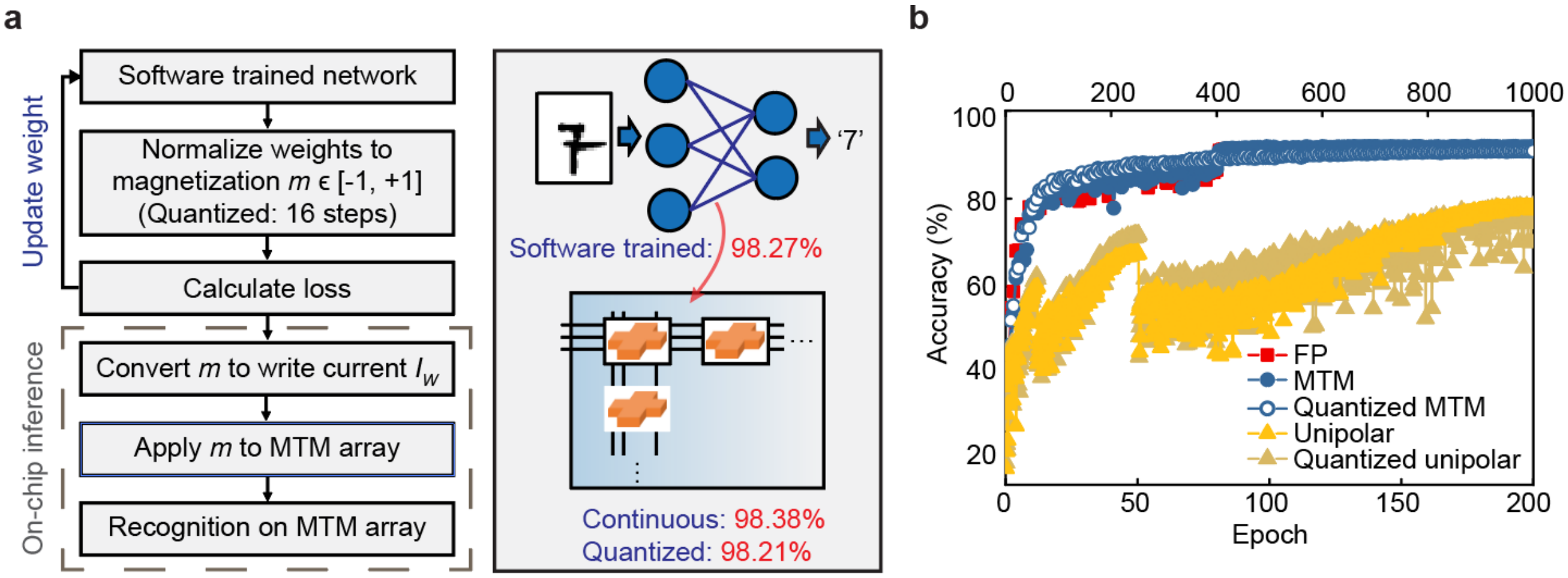

**Extended Data Fig. 7. Training scheme and CIFAR results. a**, Quantized NN training scheme. **b**, Training curves of the Floating-point software, Bipolar device, and Unipolar device ResNet-20 on CIFAR-10. Sudden jumps in the training curves happen when the learning steps change. Baseline: software; MTM: floating point; quantized MTM: 4-bit; unipolar: floating point with boost algorithm[32]; quantized unipolar: 4-bit with boost algorithm[32].

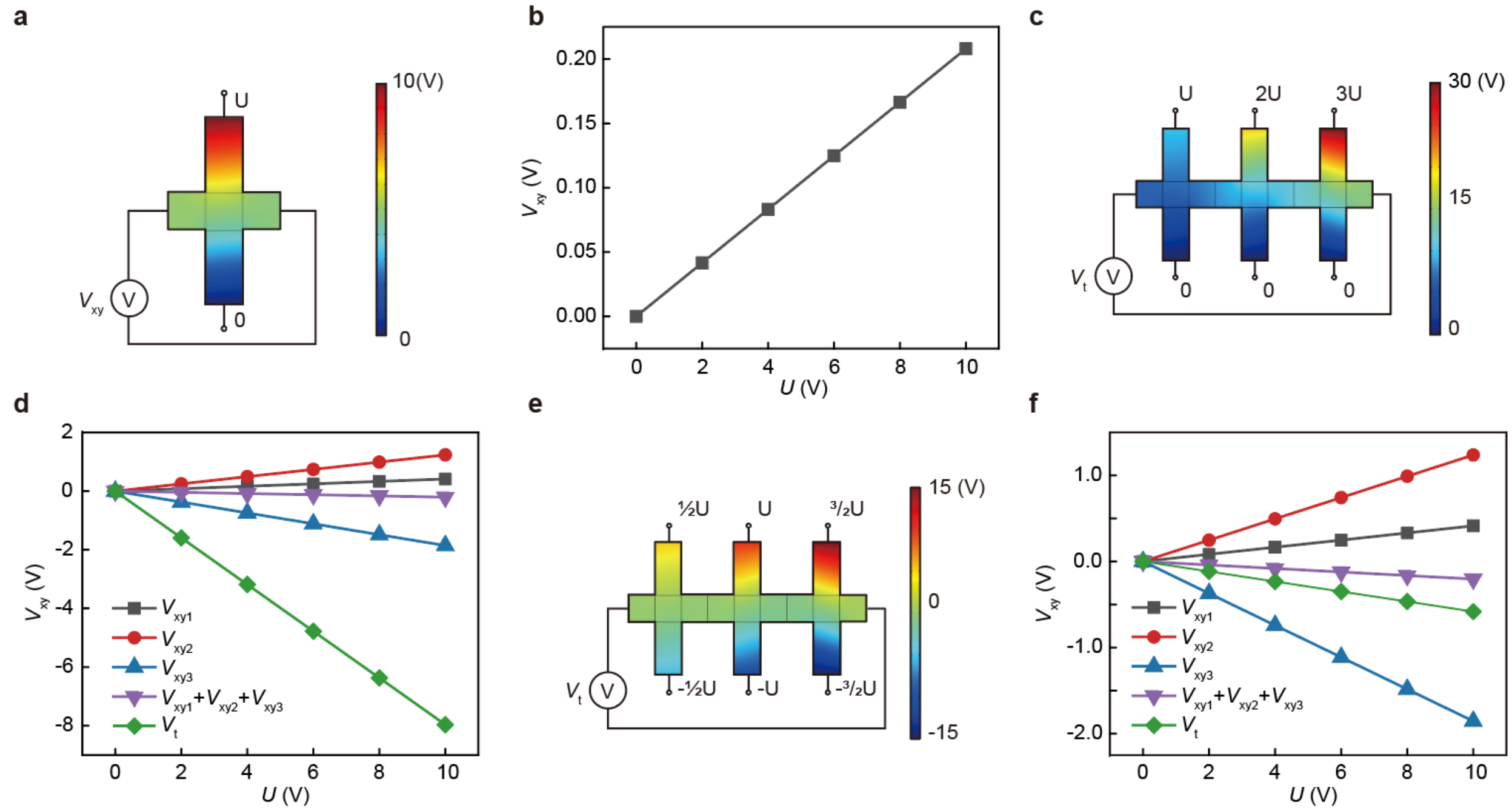

**Extended Data Fig. 8. Simulation results of the serial connection of multiple Hall channels.** **a**, Model and simulation results of a single Hall bar device. $U = 10$ V. **b**, The calculated Hall voltage $V_{xy}$ as a function of the terminal voltage. **c**, **e**, Model and simulation results of 3 Hall bar devices with Hall channel connected in serial. $U = 10$ V. **d**, **f**, The calculated Hall voltage $V_{xy}$ as a function of the terminal voltage. $V_t$ is the total Hall voltage of the 3-Hall bar device, $V_{xy1}$, $V_{xy2}$ and $V_{xy3}$ are the Hall voltage when each of the devices is connected alone.

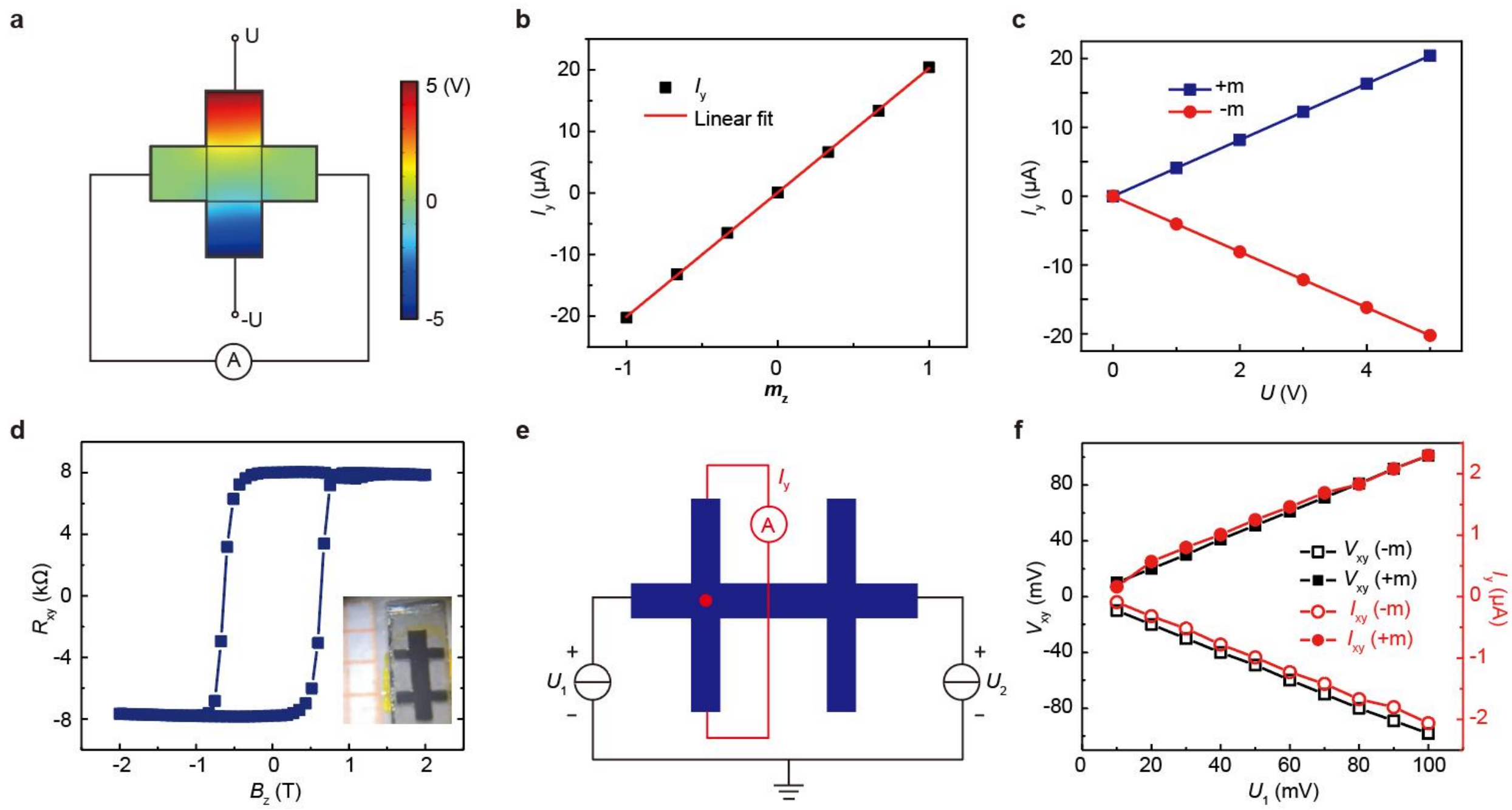

**Extended Data Fig. 9. Model and experimental verification of anomalous Hall current (AHC). a**, Model and simulation result of the single Hall bar, where the colour represents the voltage potential. $U = 5$ V. **b** and **c**, The lateral current as a function of magnetization and longitudinal voltage $U$, respectively. **d**, The hysteresis of an MTI sample D7. The inset shows a picture of the device. **e**, Experimental set-up of measuring anomalous Hall voltage (AHV) and AHC. We apply a longitudinal voltage $U_1$ and keep the voltage of the red spot to be zero by adjusting the $U_2$. **f**, AHV and AHC as a function of $U_1$.

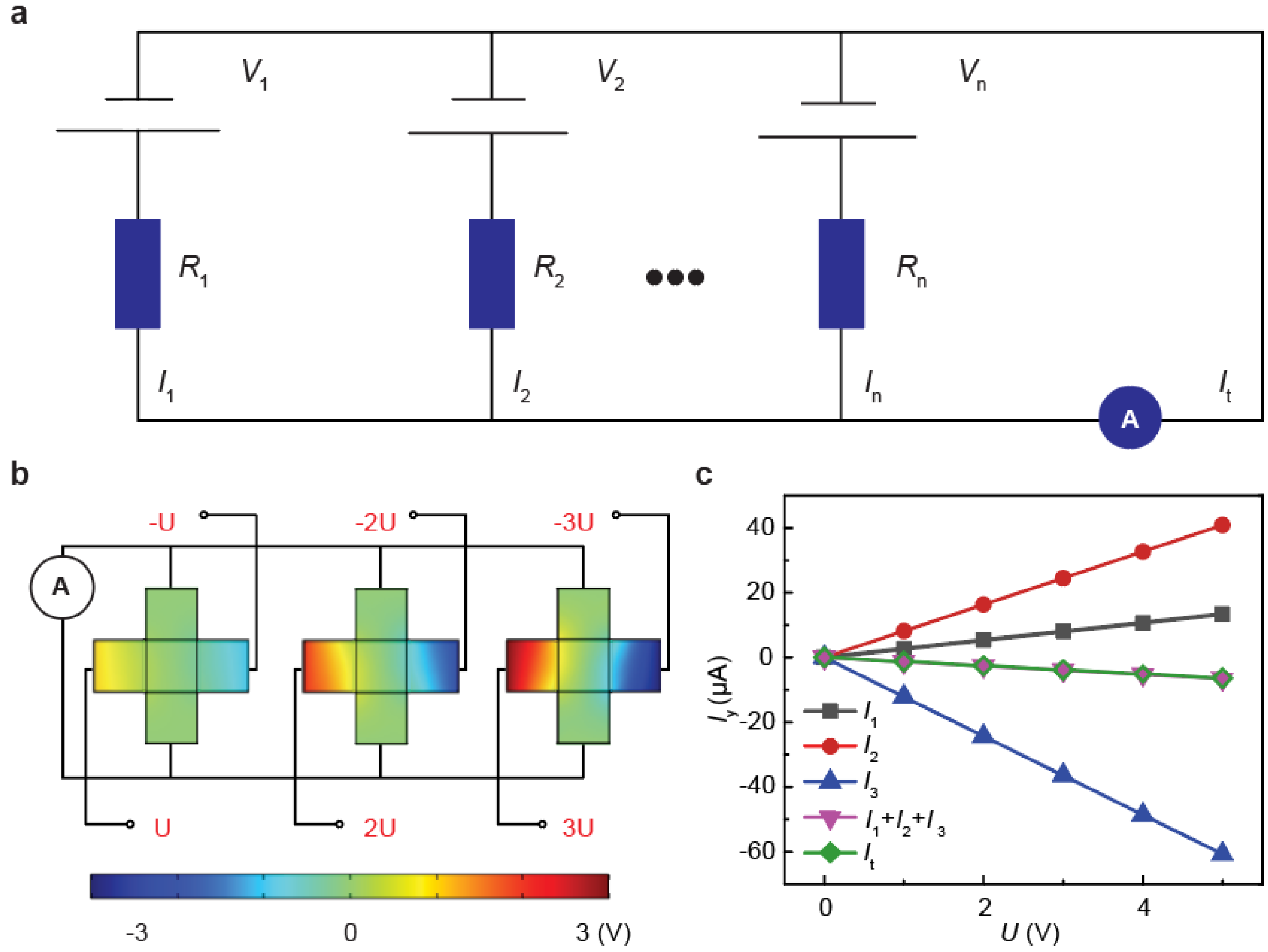

**Extended Data Fig. 10 | Equivalent circuit model of parallel connected Hall bar device and simulation verification.** **a**, Equivalent circuit model of parallel connected Hall bar devices. Each Hall bar device is treated as a voltage source with an EMF of $V_n$ and internal resistance of $R_{sy}$. **b**, Device model of three-Hall bar device with the Hall channel connect in parallel in COMSOL. **c**, Simulation verification of the Hall current summation. The plot shows the lateral current as a function of the terminal voltage, where $I_t$ is the total current output of the three Hall bar. $I_1$, $I_2$ and $I_3$ are the Hall current when each of the devices is connected alone.

**Supplementary Information for "Cryogenic in-memory computing using magnetic topological insulators"**

Yuting Liu[1,2*], Albert Lee[3*], Kun Qian[1,4*], Peng Zhang[3], Zhihua Xiao[1,5], Haoran He[3], Zheyu Ren[1,4], Shun Kong Cheung[1], Ruizi Liu[1,4], Yaoyin Li[2], Xu Zhang[1], Zichao Ma[1], Jianyuan Zhao[2], Weiwei Zhao[2], Guoqiang Yu[6], Xin Wang[7], Junwei Liu[4,8], Zhongrui Wang[9], Kang L. Wang[3], & Qiming Shao[1,4,5,8,10†]

[1]Department of Electronic and Computer Engineering, The Hong Kong University of Science and Technology, Clear Water Bay, Kowloon, Hong Kong SAR 999077, China

[2]School of Integrated Circuit, Harbin Institute of Technology, Shenzhen 518055, China.

[3]Device Research Laboratory, Department of Electrical and Computer Engineering, University of California, Los Angeles, California 90095, USA.

[4]IAS Center for Quantum Technologies, The Hong Kong University of Science and Technology, Hong Kong, China

[5]ACCESS – AI Chip Center for Emerging Smart Systems, InnoHK Centers, Hong Kong Science Park, Hong Kong, China

[6]Beijing National Laboratory for Condensed Matter Physics, Institute of Physics, University of Chinese Academy of Sciences, Chinese Academy of Sciences, Beijing, 100190 China

[7]Department of Physics, The City University of Hong Kong, Hong Kong SAR 999077, China

[8]Department of Physics, The Hong Kong University of Science and Technology, Clear Water Bay, Kowloon, Hong Kong SAR 999077, China

[9]Department of Electrical and Electronic Engineering, the University of Hong Kong, Pokfulam Road, Hong Kong SAR 999077, China

[10]Guangdong-Hong Kong-Macao Joint Laboratory for Intelligent Micro-Nano Optoelectronic Technology, The Hong Kong University of Science and Technology, Hong Kong, China

*Equal contribution

†Email: eeqshao@ust.hk

## Supplementary Note 1. Non-reciprocal resistance in MTIs

The co-existence of CES and TSS is also confirmed by the appearance of non-reciprocal resistance[1]. The non-reciprocal resistance results from the effective scattering between two edges due to presence of TSS as depicted by **Supplementary Fig. 2a**. The non-reciprocal resistance reaches maximum when applying an out-of-plane magnetic field, and the sign of the non-reciprocal resistance changes when the field direction flips. This is consistent with the change of CES from one to the other side of the device when the magnetization switches from up to down or from down to up. Note that this non-reciprocal resistance is different from the unidirectional magnetoresistance effect in MTI/TI heterostructures, which has different origins and is minimized when the magnetization is out-of-plane [2,3]. To capture the current-dependent non-reciprocal resistance, we conducted a harmonic measurement with out-of-plane magnetic field on the MTI device. Due to the chiral nature of non-reciprocal resistance, it shows opposite sign at top and bottom edges resulting in an opposite asymmetry for the second order $V_{23}^{2\omega}$and $V_{56}^{2\omega}$ shown in **Supplementary Figs. 2b** and **c**, respectively. The non-reciprocal resistance for each edge also reverses as the magnetization order flips. The non-reciprocal resistance $R_{23}^{2\omega}$and $R_{56}^{2\omega}$ has a linear dependence with current at a low current range and their values reach a large value of 400 μV at 50 μA (**Supplementary Fig. 2d**), which excludes a possible thermoelectric effect-induced voltage (assuming a thermoelectric coefficient of 2 μV/K and maximum temperature raise of 18 K)[1].

## Supplementary Note 2. Characterization of SOT and device energy efficiency

To evaluate the SOT and energy efficiency, we compare the efficiency of MTI with a control Ta (3nm)/CoFeB(1nm)/MgO(2nm)/$TaO_x$(3nm) sample and other heavy metal/ferromagnetic bilayers. The efficiency is compared in three aspects.

The first one is the SOT efficiency $\xi_{DL}$, which is defined as: $\xi_{DL} = \frac{2e}{\hbar} \cdot \frac{M_s t_m \mu_0 H_{DL}}{J_0}$. To quantify $\xi_{DL}$ of MTI, we perform a second-harmonic measurement. When the angle between the magnetic field and the x axis of the sample is 45°, the second harmonic Hall resistance $R_H^{2\omega}$ acquires a simple form as: $R_H^{2\omega} = \frac{R_s H_{DL}}{2\sqrt{2}(|H_{ext}|-H_K)}$, where $R_s$ is the saturation anomalous Hall resistance. **Supplementary Fig. 3a** depicts the typical high-field $R_H^{2\omega}$ signals produced by currents from 1 μA to 9 μA and the frequency of 181 Hz. Fitting the experimental results of $R_H^{2\omega}$ at the high field region (shown by black dashed lines in **Supplementary Fig. 3a**) yields the $H_{DL}$ strength. The extracted $\mu_0 H_{DL}$ vs current is plotted **Supplementary Fig. 3b**. By extracting the ratio between $\mu_0 H_{DL}/J_0$ and an $M_s$ of 16000 A/m[5], we obtain $\xi_{DL}$ of the MTI is about 19.2.

$\xi_{DL}$ of CoFeB sample is also characterized by the second harmonic measurement but with the magnetic field aligned with the longitudinal direction, as the anisotropy of this CoFeB sample is comparably large ($\mu_0 H_K > 500\ mT$ as shown in **Supplementary Fig. 3c**). The first harmonic Hall voltage $V_H^{\omega}$ is expected to vary with $H_{ext}$ as a cosine function or approximately a quadratic function near the extermal magnetic field $H_{ext} = 0$ with curvature $\zeta = \partial^2 V_H^{\omega}/\partial H_{ext}^2$, and the second harmonic Hall voltage $V_H^{2\omega}$ is expected to vary linearly with $H_{ext}$ with slope $\beta_L = \partial V_H^{2\omega}/\partial H_{ext}^L$. As the planar Hall effect is relatively small in CoFeB, the damping-like effective field is calculated as[9]: $H_{DL} = -\frac{2}{\zeta} \cdot \beta_L$. The first and second harmonic Hall voltages as a function of the longitudinal magnetic field at 290 K and 2 K are presented in **Supplementary Figs. 3d** and **e**. $\mu_0 H_{DL}$ as a function of injecting current is shown in **Supplementary Fig. 3f**. $\xi_{DL}$ of the CoFeB sample at 2 K and 290 K is then calculated, which only increases from 0.06 to 0.07 when the temperature decreases from 290 K to 2K. $\xi_{DL}$ of the CoFeB samples is about 300 times smaller than the MTI at 2K. Compare with MTI showing dramatic temperature dependence, the result is in agreement with our statement in the manuscript that the efficiency of normal magnetic stacks is temperature insensitive. Note that during the writing pulse, the device temperature rises to around 20 K, which can cause a significant drop of $\xi_{DL}$ as the previous research on an MTI/TI bilayer shows[10]. This highlights the importance of minimizing Joule heating for maintaining high efficiency.

The second one is normalized theoretical switching power $P_{SOT} = \rho/\xi_{DL}^2$. The resistivity of the CoFeB sample is also measured, and yields a value of 169 μΩ·cm at 290 K and 191 μΩ·cm at 2 K. We then calculate $P_{SOT} = \rho/\xi_{DL}^2$. At 290 K and 2K, they are $4.7 \times 10^{-4}$ and $3.9 \times 10^{-4}$, respectively. Compared with CoFeB, MTI is much more efficient with a $P_{SOT}$ of $3.8 \times 10^{-7}$. We note that estimations of $\xi_{DL}$ and thus $P_{SOT}$ in MTIs are subject to considerable uncertainty due to variations in measurement protocols and samples [10,11]. Therefore, we propose a more experiment-oriented aspect below using the experimentally obtained switching current density.

The third one is normalized experimental switching power $P_{SW} = \rho J_{sw}^2$. As the second aspect is only working for single domain switching, whereas the switching in large devices is done through domain nucleation and domain wall propagation. To compare the experimental energy efficiency of both devices at deep cryogenic temperature, we also perform out-of-plane field switching and SOT switching experiments for the CoFeB sample at 290 K and 2 K as shown in **Supplementary Figs. 3g** and **3h**, respectively. To achieve a higher switching ratio, the in-plane magnetic field during SOT switching increases from 30 mT to 120 mT from 290 K to 2 K. The switching ratio is 80% at 290 K and 15% at 2 K. We obtain the critical current density for the CoFeB sample is $1.7 \times 10^7$ A/cm$^2$ at 290 K and $2 \times 10^7$ A/cm$^2$ at 2K. The normalized experimental switching power of CoFeB sample, calculated as $P_{SW} = \rho J_{sw}^2$, yields a value of $4.9 \times 10^{16}$ W/m$^3$ at 290 K and $7.6 \times 10^{16}$ W/m$^3$ at 2K. Compared with the CoFeB sample, $P_{SW}$ of MTI reaches a low value of $2.5 \times 10^{15}$ W/m$^3$. The experimental switching power is also much lower than CoFeB. A full efficiency comparison of MTI with other heavy metals is summarized in **Supplementary Table 1**. The comparison suggests that MTI possesses both superior SOT efficiency and switching power efficiency at 2 K.

### Supplementary Note 3. Mechanism of SOT switching in MTIs

In the past experiments, we have shown that the Cr-BST/BST gives rise to a giant spin-orbit torque and the origin of the large spin-orbit torque efficiency is due to the spin-momentum locking of topological surface states by tuning the relative ratio between surface states and bulk states in a single Cr-BST layer using a gate voltage[5,6]. Note that while the Cr-BST is nominally uniformly doped, the different dielectric environment on the top ($AlO_x$ capping layer) and bottom (GaAs substrate) produces a net spin current from the top and bottom surface states. As a result, we show that the magnetic order and its associated CES in the Cr-BST can be switched by the in-film plane current injection (**Fig. 1d**). Note that to break symmetry for switching magnetization along the out-of-film plane direction, we apply a small assistance field of 30 mT.

The critical switching current density is about $4.2 \times 10^5$ A/cm$^2$ for device D1 (**Fig. 1d**), which lifts the sample temperature to about 20 K during SOT switching (**Extended Data Fig. 1d**). As this temperature is still below $T_c$ by a big margin, MTI remains ferromagnetic (**Extended Data Fig. 2a**) even if the reading current pulses reaches $4.2 \times 10^5$ A/cm$^2$. For another device D6, the critical switching current is about 1.25 $\times$ $10^5$ A/cm$^2$ (**Supplementary Fig. 4c**), which lifts the temperature to about 13 K during switching (**Supplementary Fig. 4d**). So, for different devices, the device temperatures during switching are different, but they are all below $T_c$ by a big margin. Hence the current induced switching is led by SOTs.

To directly evaluate the nature of the switching behavior, $R_{xy}$ measured by reading and writing pulses during current-induced switching is presented in **Supplementary Fig. 5**. The current switching experiment is measured by a two-step scheme, where a 2 ms writing pulse (up to 1 mA, $8.3 \times 10^5$ A/cm$^2$) is sent first and then followed by a reading pulse (10 μA, $8.3 \times 10^3$ A/cm$^2$) after 100 ms. $R_{xy}$ recorded by the writing pulse is getting smaller when the writing current is getting larger, while $R_{xy}$ recorded by the reading pulse remains almost unchanged after switching. This behavior suggests the writing pulse indeed heats up the sample. But the finite difference of $R_{xy}$ at $\pm 8.3 \times 10^5$ A/cm$^2$ indicates that the sample is still below Cuire temperature and ferromagnetic (**Supplementary Fig. 5b**). Meanwhile, MTI is already cooled down before the reading pulse arrives as $R_{xy}$ recorded by the reading pulse remains almost unchanged after switching. The domain pattern during current switching can't be a single domain structure, as the 750 Ω current switchable range is much smaller than a $R_{xy}$ of 3800 Ω obtained by the field switching experiment with a large current density of $4.2 \times 10^5$ A/cm$^2$ (**Extended Data Fig. 2a**). Hence, we expect a formation of multi-state domain structure. This argument can also be evident by comparing $R_{xx}$ during current and field switching. $R_{xx}$ recorded by the reading pulse yields a value of around 13.9 kΩ for the current switching experiment (**Supplementary Fig. 6a**), which corresponds to $R_{xx}$ of the multi-state domain state (around coercive field of 0.1 T) for the field switching experiment (**Extended Data Fig. 1b**).

Overall, we conclude that the current switching of our MTI is driven by SOT, and the current-induced heating effect results in a partial switching and a formation of multi-domain states resulting in a smaller current-tunable AHR range. Nevertheless, the tunable AHR range is two orders of magnitude larger than that of a technology-relevant ferromagnet, CoFeB (around several Ohms)[7,8], which makes the readout using AHR feasible. The summarized device properties of D1-D4, and D6 are provided in **Supplementary Table 2**.

**Supplementary Note 4. Memristive switching behaviors in MTIs**

**Fig. 2a** illustrates the measurement setup. Four devices are characterized which will be used for the classification of the IRIS dataset. Each device can be independently addressed via multiplexers and demultiplexers commanded by a microcontroller unit (MCU). The field-induced switching of the four MTI devices is shown in **Fig. 1b** and the current-induced SOT switching in **Fig. 1d**. The devices show uniform and consistent field-induced and current-induced switching properties. For the field switching, all devices have a giant AHR of about 11 kΩ and a coercivity of 100 mT. For current-induced switching, all devices can be switched by a current of 0.5 mA (current density of $4.2 \times 10^5$A/cm$^2$) and have an AHR of -600 Ω to 600 Ω. Compared with normal HM/FM structures, the switching current density is more than one order of magnitude lower, and the AHR is more than two orders of magnitude higher.

We then characterize the switching curve (e.g., write current vs. AHR) of the device. In this experiment, we reset the resistance of the memristor via a -1 mA ($-8.3 \times 10^5$ A/cm$^2$) pulse before each write pulse, followed by applying a read pulse of 30 μA ($2.5 \times 10^4$ A/cm$^2$) to readout the Hall resistance. The switching curves acquired from 50 trials are shown in **Fig. 2b**. It should be noted that although 12 resistance levels are shown, all AHR values within the AHR range can be achieved. **Supplementary Fig. 7a** shows the write distribution of the MTI devices, presented as the difference between the written AHR and the mean written AHR of 50 trials for each of the 12 resistance levels above. The maximum write variation is less than 25 Ω and the standard deviation error is 7.6 Ω, which corresponds to about 1.9% of the writing AHR range (-200 Ω to 200 Ω). This write variation is substantially smaller than other memristor devices that operate on the stochastic formation and rupture of conducting channels.

The reading distribution of the MTI device is characterized by first applying a write pulse followed by 90 consecutive read pulses. We first test whether the AHR is stable or not across different read pulse amplitudes from 1 μA ($8.3 \times 10^2$ A/cm$^2$) to 56 μA ($4.6 \times 10^4$ A/cm$^2$), which is necessary for analog multiplication using the AHR neural network. It can be observed in **Fig. 2c** that the measured Hall voltage is very stable over repeated reading attempts, suggesting a small read noise of the device and low disturbance. The distribution of reading noise, defined as the difference between the measured AHR and the mean measured AHR of the 90 read operations, is shown in **Supplementary Fig. 7b**. We observe a maximum reading noise of 1.5 % and a standard deviation as small as 0.37%.

In addition to the above noises, the thermal effect on CES introduces a reading discrepancy as AHR reduces when the reading current increases (see **Extended Data Fig. 2**), which creates a nonlinearity during inference that potentially impacts the vector-matrix multiplication (VMM) operation. **Supplementary Fig. 7c** shows AHR as a function of reading current at different magnetization states. The Hall resistance for the same magnetization state changes dramatically when the reading current density is below $8.3 \times 10^2$ A/cm$^2$, while it remains relatively stable when the reading current density is above $8.3 \times 10^3$ A/cm$^2$. Hence, the AHR measurement presented in the main manuscript is measured by a reading current density of $8.3 \times 10^3$ A/cm$^2$. Meanwhile, to evaluate the impact of reading discrepancy on the inference test, the data from $1.6 \times 10^4$ A/cm$^2$ A/cm$^2$ to $3.2 \times 10^4$ A/cm$^2$ are collected for making the statistics of reading discrepancy noise. The Hall resistance measured by $2.5 \times 10^4$ A/cm$^2$ pulses is set as the reference, and the reading error is defined as the difference between the reference value and the resistance measured by other pulse amplitude. As shown in **Supplementary Fig. 7d**, there is at most a 5% reading discrepancy to the reference resistance value in the interesting range and the standard deviation error is about 2%.

**Supplementary Note 5. Large-scale dataset simulation**

To further evaluate the MTI device and show the importance of bipolar weights, we demonstrate image recognition simulation on larger neural networks using the MNIST and CIFAR datasets. For the MNIST dataset, the neural network has two layers, the first hidden layer with 150 neurons and the second classification layer with 10 neurons as shown in **Fig. 3a**. The algorithm is depicted in **Extended Data Fig. 7a.** The floating-point and bipolar networks are trained using stochastic gradient descent for 200 epochs with a batch size of 128, initial learning rate (lr) of $10^{-3}$, and a cosine learning rate schedule[12]. A L2 weight loss of $10^{-4}$ encourages the weights to be near 0. The unipolar network is trained for 1000 epochs with a batch size of 128, weight decay of $10^{-4}$, an initial learning rate of $10^{-2}$, and a cosine lr schedule[13]. Additionally, we implement the algorithm in Ref. [13] to improve its performance. A summary of training parameters is shown in **Supplementary Table 3.**

We compare the performance of the neural network with different device models (floating-point neural network, MTM neural network, and unipolar neural network). Memristance of the floating-point neural network can take any real values without limitation, while that of the MTM neural network is bounded to be -800 Ω to 800 Ω and subject to a 4% (2% for writing and 2% for reading) Gaussian noise. The memristance of the unipolar neural network only takes positive resistance values from 1000 Ω to 3000 Ω. As shown in **Fig. 3b**, the software-trained neural network FP achieves an accuracy of 98.27% and MTM neural network achieves a final accuracy of 98.38%, in contrast to the unipolar neural network achieving a final accuracy of 94.26%. **Fig. 3c** presents the normalized weight of each neural network after training. The weight patterns of different synapses of MTM neural network almost replicate the result of the floating-point neural network indicating the in-situ training of our MTM with bipolar weights parallels that of the software. For the unipolar neural network, however, the performance is bounded by the limited range of weights, such as in classifying characters ”1” and ”2”, owing to lacking negative resistances. Meanwhile, we consider a more practical case in which MTM states are quantized. For this design, AHRs are quantized to the nearest value of 16 steps between [-1, +1] (normalized) before calculating the write current. The quantized MTM network shows a less than 1% accuracy drop to 98.21%. This result confirms the superior performance of MTM and the crucial role of bipolar weights in implementing deep learning models.

For the CIFAR-10 dataset, the neural network is the ResNet-20 with three residual blocks for a total number of 20 layers. The bipolar network is trained via SGD optimizer for 200 epochs with batch size of 128, weight decay of $10^{-5}$, and an initial learning rate of $10^{-3}$ with a multi step lr decay schedule. The unipolar network parameters were optimized via Bayesian optimization and trained for 1000 epochs with a weight decay of $2 \times 10^{-4}$ and an initial learning rate of $7 \times 10^{-2}$ with the cosine lr schedule. A summary of training parameters is shown in **Supplementary Table 3.** As shown in the **Extended Data Fig. 7b**, the FP neural network achieves an accuracy of 91.6% and MTM neural network achieves an accuracy of 91.9%, while the unipolar neural network performs a much lower accuracy of 78.1%. The quantized neural networks show similar accuracies with a difference of less than 1%.

**Supplementary Note 6. MTM neural network design**

To leverage the efficiency advantage of computing with a large-scale MTM neural network hardware, we design a novel circuitry to overcome the challenges mentioned in the Methods section ('Design challenges for Hall effect-based neural network**')**. To solve the issue that the Hall signals don't sum correctly, we read anomalous Hall current (AHC) instead of AHV for Hall signal summation. The Hall bar devices are connected in parallel during VMM operation while being isolated by transistors during read/write. The validation of this current-mode readout scheme is presented in the Methods section ('Verification of Hall current mode reading'). Another issue for the MTM neural network is that the AHE signal could be too small when the device size scales down. The reduced read current and device dimensions result in a significantly smaller readout signal. To overcome this, we flow the read current perpendicular to the external field direction and collect AHC along the field direction. The magnetization states would not be changed by the SOT effect due to symmetry[8]. The read disturbance would be much lower, hence allowing us to increase the read current for a higher AHC signal.

The schematic of the MTM neural network is shown in **Supplementary Fig. 8a**. Three transistors are introduced to each MTM to form a memory cell. Transistor $T_T$, controlled by WWL, connects the top node T to bus SL. Transistors $T_L$ and $T_R$, both controlled by RWL, connect the lateral nodes L and R to bus BL and BLB, respectively. The bottom node B is connected to SLB directly. In the array, cells in the same row share common WWL and BLB buses, while cells of the same column share the same SL, SLB, and BL. The unconventional design of running SL and SLB perpendicular to each other is necessary to enable neural network and read operations within the same array. **Supplementary Fig. 8b** illustrates the waveforms of the MTM array during memory and VMM operations. During a memory write, the WWL of the selected row is activated, while the rest WWLs and all RWLs are grounded. SL and SLB are biased to the write conditions, e.g., $V_{SLi} = V_{w,i}$ , $V_{SLBi} = V_{SS}$. During a memory read, both the WWL and the RWL of the selected row are activated, all BLs and BLBs are biased at the read voltage with the center node of the AHR virtually grounded, e.g., $V_{BLi,\,BLBi} = V_{VGND} \pm V_{RBL}$, and all SLs and SLBs are clamped to the lateral read voltage, e.g., $V_{SLj,SLBj} = V_{VGND} \pm V_{RSL}$. The AHC of the selected memory cells will be accumulated on each SL as $I_{SL,j}$. The virtual ground VGND design avoids electrical current between the horizontal and the lateral channels that disrupts the readout. The schematic description of the write and read operations are depicted in **Supplementary Fig. 9**. During a VMM operation, all WWLs and RWLs are turned on. Voltages corresponding to the neural network inputs are applied to BL and BLB, e.g., $V_{BLi,BLBi} = V_{VGND} \pm V_{in,i}$. The SLs and SLBs are clamped to the lateral read voltage in reverse, e.g., $V_{SLj,SLBj} = V_{VGND} \pm V_{RSL}$. The neural network output currents are read from each SLB as $I_{SLB,i}$. The schematic description of the VMM operation is shown in **Supplementary Fig. 8c**.

**Supplementary Note 7. Circuit simulation**

We simulate the MTM neural network on the commercial design platform Cadence using a foundry process. The MTM model is implemented as a 4-component conductance-transconductance model with respect to the center node of the MTM. The simulation result of a 3-input VMM operation is presented in **Supplementary Fig. 8d**. Three weights are stored as the magnetization of three MTMs [$m_1$, $m_2$, $m_3$] = [+1, +0, -1], and the VMM operation is conducted with different input values [$V_{in,1}$, $V_{in,2}$, $V_{in,3}$] = [+0.1V, +0.1V, +0.1V], [+0.1V, +0V, -0.1V], and [-0.1V, +0V, +0.1V]. The SLB currents ($I_{SLB}$ = 21 μA,50 μA, and -10 μA) are almost perfectly linear to the multiply-and-accumulate results (0, +2, -2). The small offset is caused by the nonzero resistance of the access transistors and parasitic along each bus, which reduces the voltage bias on the MTMs. This result further confirms the feasibility of using AHC for VMM in AHR neural networks.

Performance comparison of the MTM device, regular HM/FM Hall device, MRAM device, and tensor processing unit (TPU) is presented in **Supplementary Table 4.** The dimensions of the Hall devices are 50 nm×50 nm and 200 nm×200 nm in the cross section area while the MRAM devices are 50 nm in diameter. To calculate the energy for CIFAR-10, we first extracted the capacitance on the WLs, BLs, and SLs using parasitic extraction (PEX) on the layout of a 3×3 dummy-cell mini-array. The capacitance on the BLs and SLs was about 0.2 fF/cell and that of the WLs was about 0.4 fF/cell. The charge/discharge energy of the BLs, SLs, and WLs can be obtained as E = $CV_{\mathrm{DD}}V_{\mathrm{Swing}}$, where $V_{\mathrm{DD}} = 0.85\ V$ is the supply voltage and $V_{\mathrm{Swing}}$ are the voltages on the BLs, SLs, and WLs during operation. For the STT-MRAM, read and neural network operations use a BL voltage of 100 mV and write a voltage of 1.5 V[14]. For our MTM Hall NN, read and neural network operations use BL and SL voltages of 80 mV, and write uses an SL voltage of 80 mV for 50 nm device and 300 mV for 200 nm device, as derived from the experiments in this work. For the FM Hall NN, write uses a voltage of 50 mV and read of 70 mV. All WLs are driven to $V_{\mathrm{DD}}$. The dissipation energy can be obtained as IVT, where I is the current flowing through the memory cell, V is the supply voltage, and T is the cycle time. The NN operation has its own independent supply. Cell current is obtained by dividing the read/write voltage by the device resistance (MRAM:4 kΩ/11.2 kΩ, MTM-Hall:31 kΩ, and FM-Hall: 850 Ω in either direction), and the cycle time is set to 5 ns for R/W (100 MHz) and 2 ns (250 MHz) for NN operation. The decoder energy was extracted using simulation of the decoding path built upon 3-to-8 and 4-to-16 pre-decoders, resulting in the energy of ~1 fJ/bit. The sense amplifier energy was extracted using a simulation of a latch-type sense amplifier and consumed up to 4 fJ/bit. The write driver energy was extracted from a FO4-sized buffer chain that consumed ~10 fJ/bit. During neural network operation, an ADC energy of 1.9 pJ is consumed for each row. The total energy for read, write, and NN operations includes the activation of each bus, its drivers, controlling circuitry, and the DC current through memristors. Finally, we consider the number of write/read cycles and arrays required to carry out each operation: In MRAM NNs, write, read, and NN operation require activation of two arrays. For Hall NN, only one array is active, but a reset operation is necessary before each write operation. The simulation parameters are summarized in **Supplementary Table 5**. Compared with the MRAM neural network, the MTM NN features a 86% lower write energy and a 11× higher TOPS/W. Compared to FM-Hall neural network, the MTM NN features a 56% reduction in write energy and a 5× higher TOPS/W. The improvement mainly comes from (a) a CES-induced large AHE signal, resulting in a low read current (b) TSS-induced low write current, and (c) the capability to represent both positive and negative weights using CES-based giant and tunable AHE on a single device.

**Supplementary Note 8. Analysis of two-terminal resistance in MTI devices**

Here, we analyze the two-terminal resistance ($R_{2T}$) that affects the device performance. The operation principles of our MTI devices require the co-existence of TSS and CES, which means that devices need to operate above the quantum anomalous Hall effect temperature. In this sense, the two-terminal resistance can be much smaller than 25.8 kΩ, which is the case for CES only, if the aspect ratio of the device is designed properly. We measure temperature dependent four-terminal resistance $R_{23,14}$ (**Supplementary Fig. 1b**) and two-terminal resistance $R_{14}$ (**Supplementary Fig. 10b)** at 1 T with a reading current of 10 μA ($8.3\times10^3$ A/cm$^2$). The distance between 1 and 4 electrodes is 40 μm and the distance between centers of 2 and 3 electrodes is 13 μm (**Supplementary Fig. 10a**). On the one hand, $R_{14}\neq25.8$ kΩ means that our device is not in the ideal QAHE regime. On the other hand, the inconsistency of two-terminal resistance values ($R_{inc}$) at 1 T using $R_{14}$- $R_{23,14}\times$ (40/13) as a function of the temperature (**Supplementary Fig. 10c**) is large and temperature sensitive in our MTIs, where 40/13 is a geometric coefficient. It turns out that the $R_{inc}$ includes the contributions from CES-induced $R_{inc,S}$ due to the transverse transport (nonzero $\sigma_{xy}$) and contact resistance from our further analysis. We can simulate the $R_{inc,S}$ by numerically solving the Laplace equation with a measured conductance matrix consisting of $\sigma_{xx}$ and $\sigma_{xy}$. Here is the procedure: First, the electric potential φ of the channel of the Hall bar satisfies Laplace's equation $\nabla^2\varphi = 0$. The simulation also obeys Ohm's law J=σE, where E=∇φ, and conservation of charge $\nabla\cdot \mathrm{J} = 0$. A DC source I is applied to the left boundary and the right boundary is electrical grounded. Top and bottom edges are insulating as $\hat{n}\cdot J = 0$. Then we can get the electric potential distribution (**Supplementary Fig. 10d**) and calculate the simulated $R_{xx,S} = V_{xx}/I$ and $R_{2T,S} = V_{2T}/I$. In the end, we confirm that the measured and simulated $R_{xx}$ are consistent (**Supplementary Fig. 10e**). Then, we observe the trends for the measured $R_{2T,E}$ and simulated $R_{2T,S}$ are very similar, i.e., increasing with the decreasing temperature (**Supplementary Fig. 10f**). We also show the temperature dependence of $R_{inc,S}$, from which we can see that the increasing trend (from 0.042 kΩ to 2.4 kΩ) is due to the transverse transport induced by the CES. Lastly, we see that the estimated contact resistance $R_c$ is fluctuating between 8.5-9.2 kΩ, which is much less temperature-dependent compared with the $R_{inc,S}$.

To further confirm the importance of the CES in the $R_{inc}$, we reduce the current density to minimize the heating and thus increase the CES contribution. We use current of 1 nA (0.833 A/cm$^2$). The results are shown in **Supplementary Figs. 10g-h**. We can see that the $R_{inc,S}$ increases from nearly zero to 4 kΩ, while the $R_c$ fluctuates between 8.1-9.8 kΩ due to a large measurement noise at such a low current level.

In short, we show that the inconsistency of the two-terminal resistance values is not solely from the contact resistance and the role of the CES contribution is essential and defines the trend for the temperature dependence. For the future work, the contact resistance can be reduced to a few ohms by optimizing the MTI/metal electrode structures, such as using a comb-like structure for the MTI layer[4].

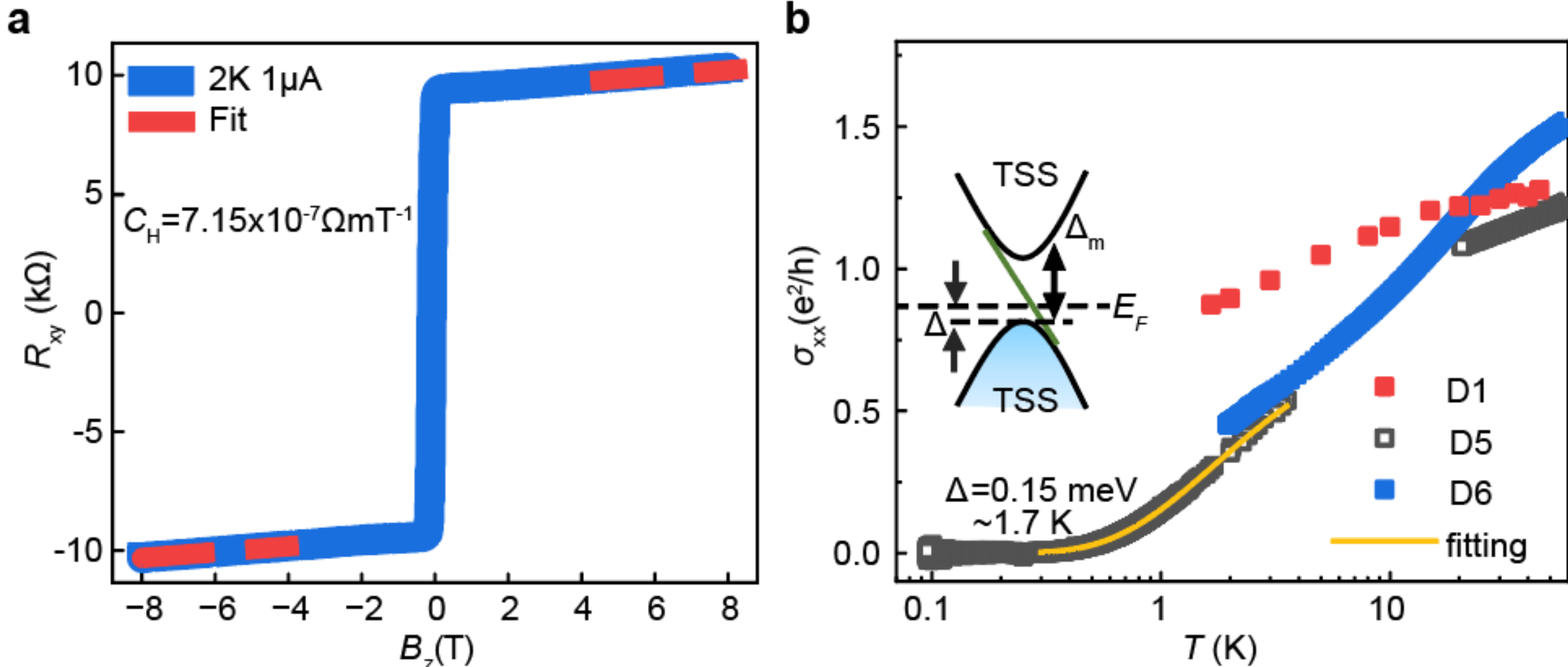

**Supplementary Figure 1**. **Estimation of the Fermi level**. **a,** The Hall effect in D1**.** The blue curve is measured with a 1 μA ($8.3\times10^{2}$ A/cm$^{2}$) current at 2 K, and the red dash line is the linear fit. Positive and small slope indicates that the sample is slightly p-doped and the Fermi level is close to the surface valence band. **b,** Temperature dependence of longitudinal conductance $\sigma_{xx}$ in devices D1, D5 and D6, respectively. A fitting of $\sigma_{xx}$ using the single–activation gap Arrhenius equation of D5 at 0.3-4 K shows the TSS activation energy gap of 0.15 meV. The inset shows a schematics of energy band structure of TSS and CES.

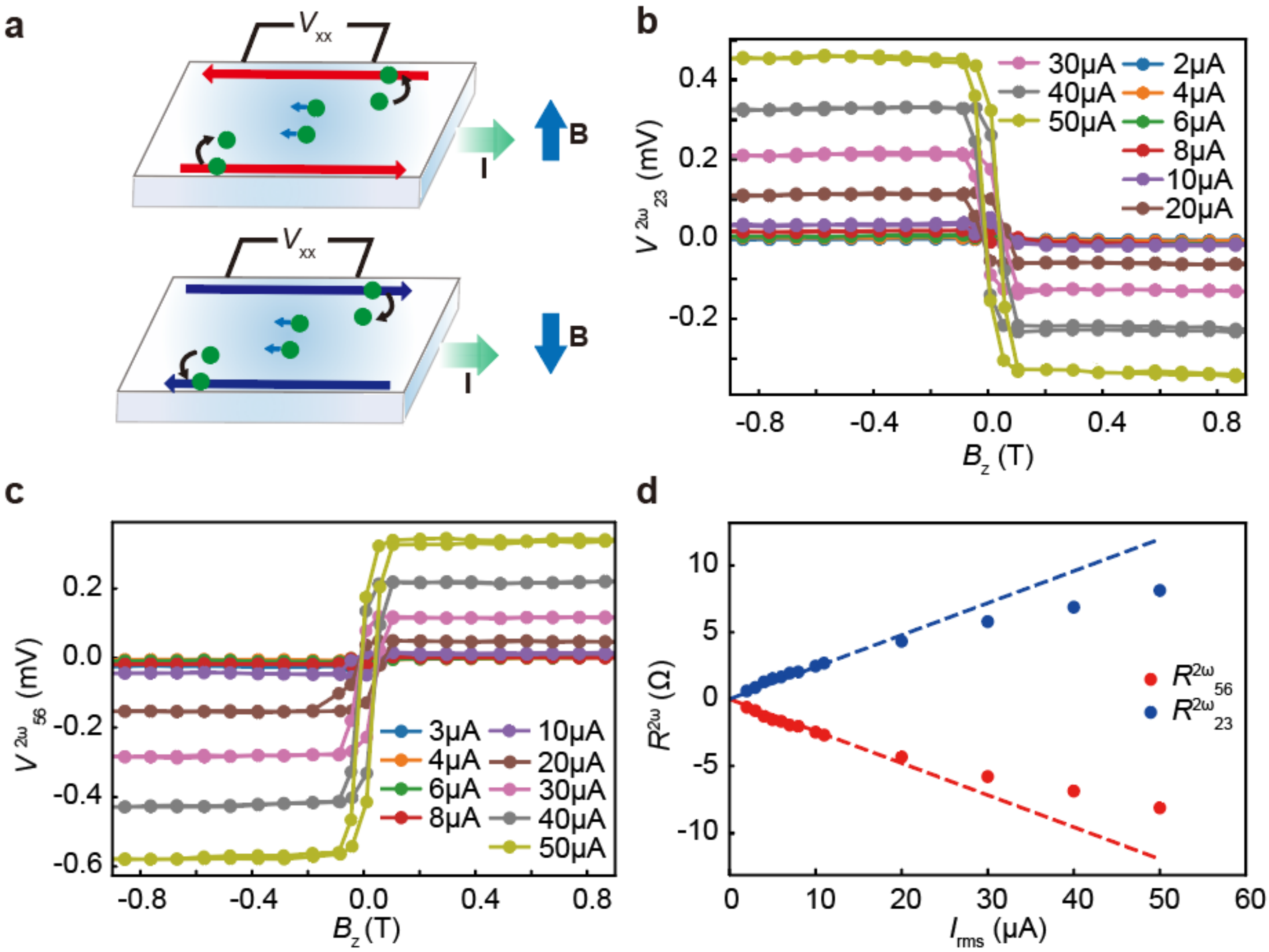

**Supplementary Figure 2**. **Second harmonic longitudinal resistance of the MTI device D1. a,** Schematic of the non-reciprocal resistance. **b-c,** Field dependence of second Harmonic longitudinal voltage with different current. **d,** Comparison of second harmonic longitudinal resistance between top and bottom connects. The data is acquired at 2 K, and the frequency of the AC current is 17 Hz.

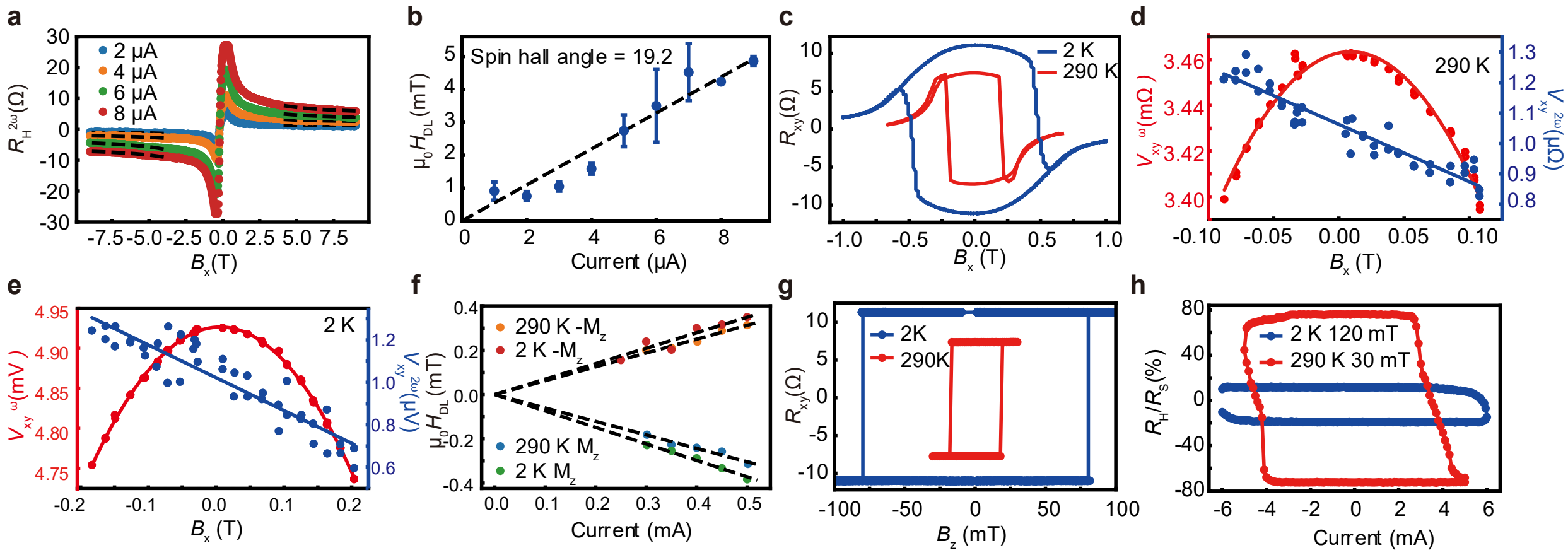

**Supplementary Figure 3**. **Characterization of SOT efficiency of MTI and CoFeB magnetic stack**. **a,** Dependence of the second harmonic Hall resistance of MTI device D1 on the magnetic field with different injecting current at 2 K. The dashed line denotes the fitting. **b,** The relationship between the damping-like field $\mu_0 H_{DL}$ and the injecting current for the MTI sample. **c,** The dependence of Hall resistance on the in-plane magnetic field at 2 K and 290 K for the CoFeB magnetic stack. **d** and **e,** First and second Hall voltage of the CoFeB sample with respect to the magnetic field at 290 K and 2 K. **f,** The relationship between the $\mu_0 H_{DL}$ and the injecting current for the CoFeB sample at 2 K and 290 K. The dashed line represents the fitting. **g,** The perpendicular magnetic field switching of the CoFeB sample at 2 K and 290 K. **h,** The current switching of the CoFeB sample at 2 K and 290 K. A 120 mT (30mT) in-plane field is applied along the x-axis of the sample at 2 K (290 K) to assist the switching.

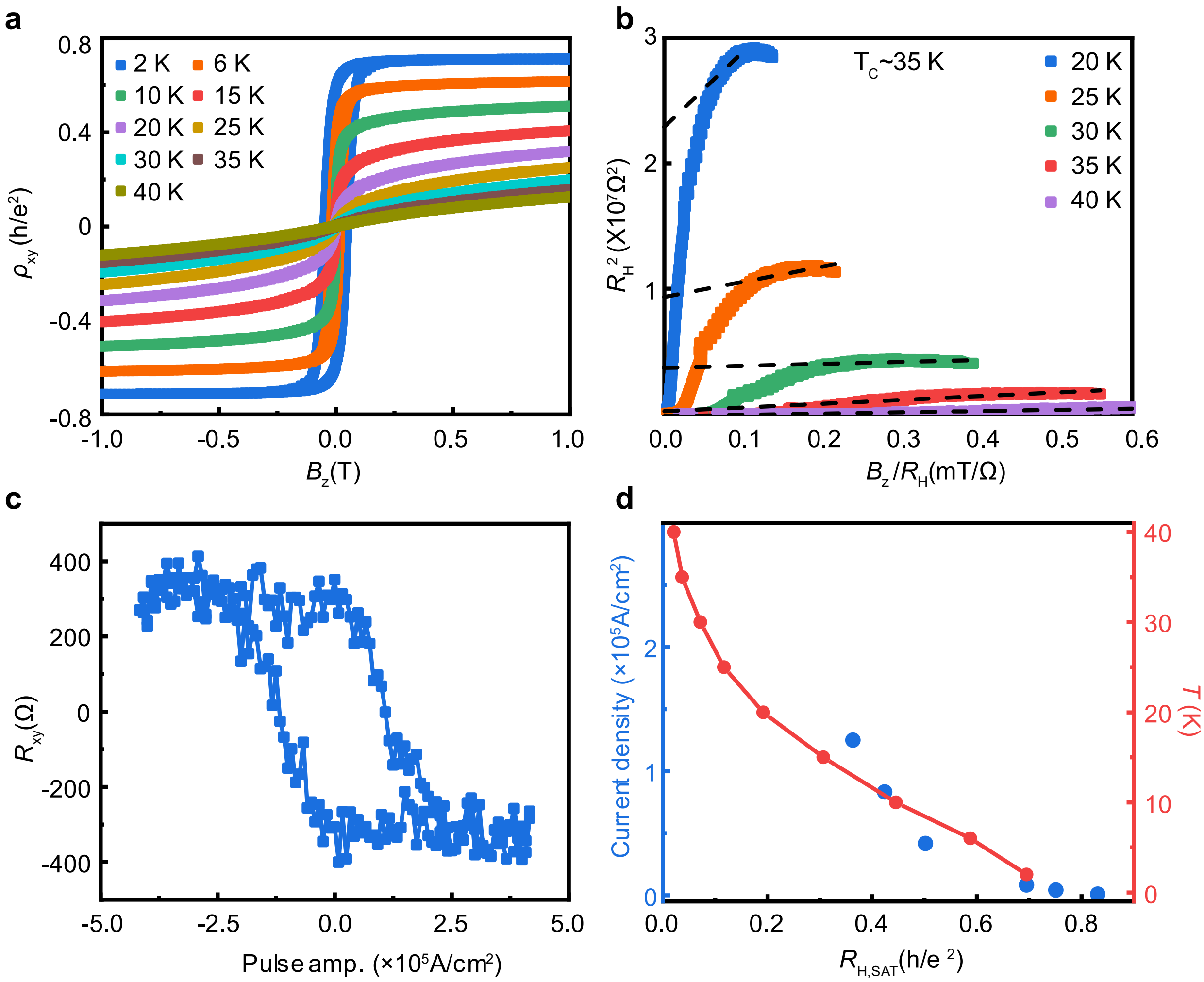

**Supplementary Figure 4**. **Magnetic properties and current-induced switching of D6. a,** Anomalous Hall resistance as a function of out-of-plane magnetic field at different temperatures measured by 8333 A/cm$^2$ pulse**. b,** Arrott plot for the MTI device, where the Curie temperature is estimated to be 35K. **c,** The pulse write current-induced switching with -15mT in-plane field at 2K. **d,** The relationship between current density and saturated AHR, and the relationship between temperature and saturated AHR.

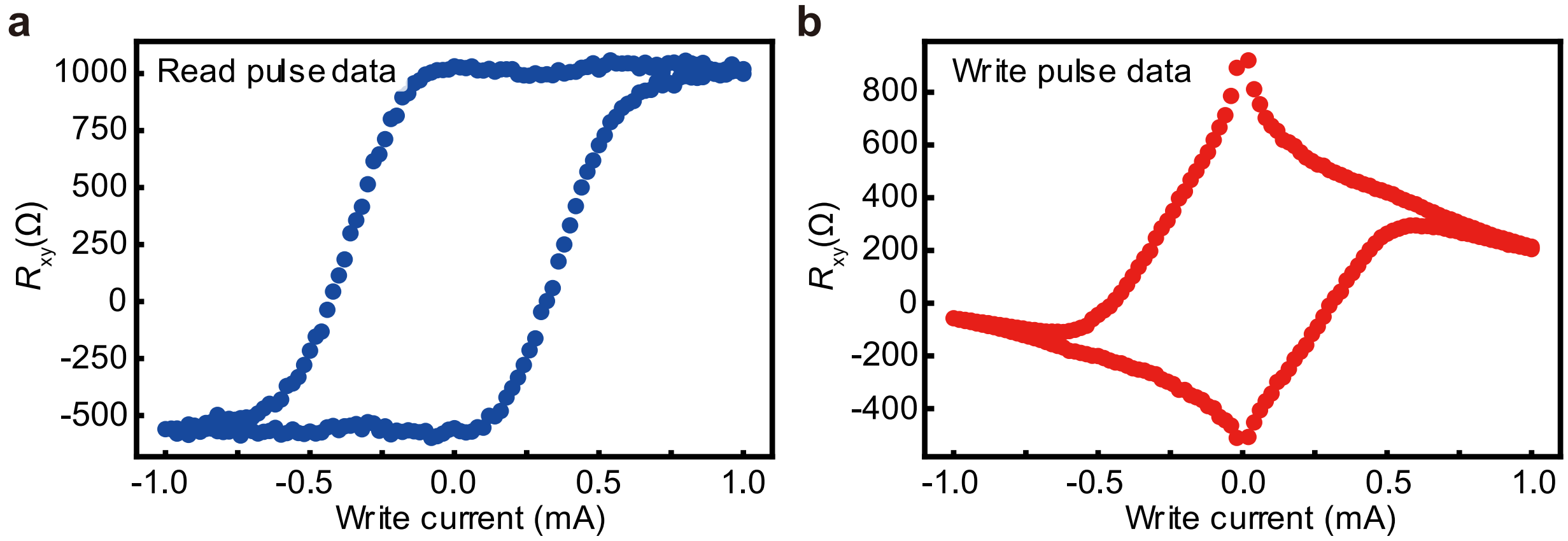

**Supplementary Figure 5**. **Values of $R_{xy}$ during read and write pulses for D1**. **a,** $R_{xy}$ recorded by reading pulses during SOT switching. **b,** $R_{xy}$ recorded by writing pulses during SOT switching.

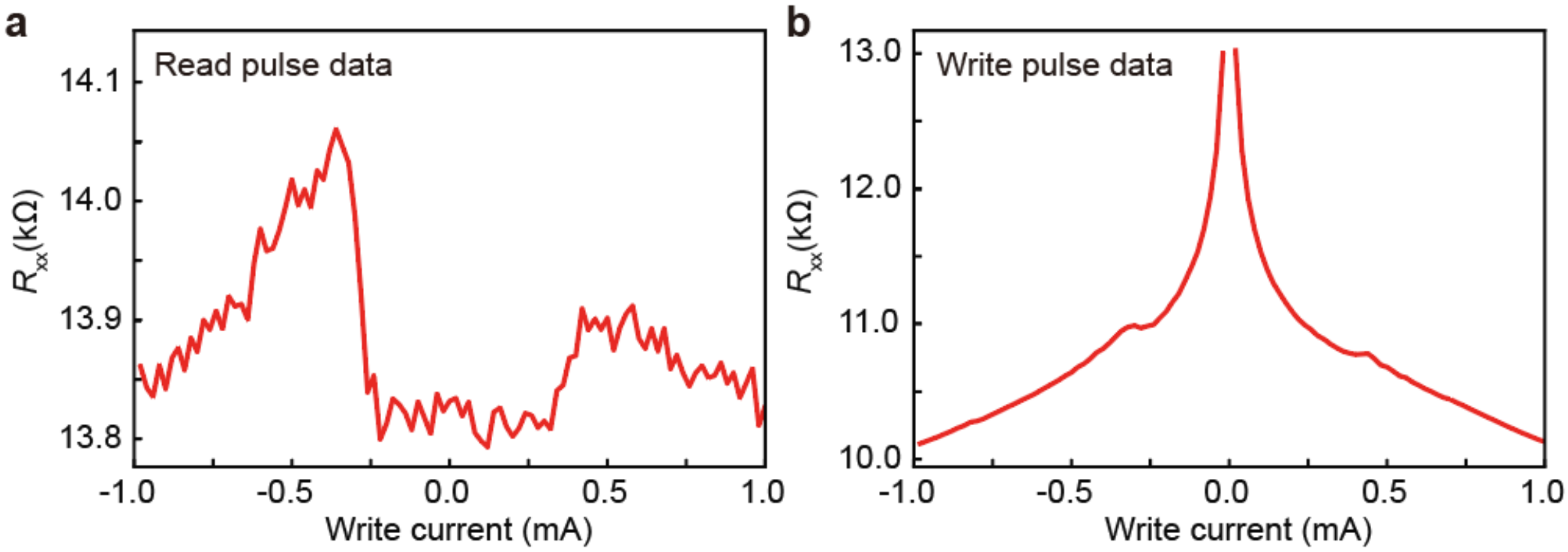

**Supplementary Figure 6**. **Values of $R_{xx}$ during read and write pulses for D1**. **a,** $R_{xx}$ recorded by reading pulses during SOT switching. **b,** $R_{xx}$ recorded by writing pulses during SOT switching.

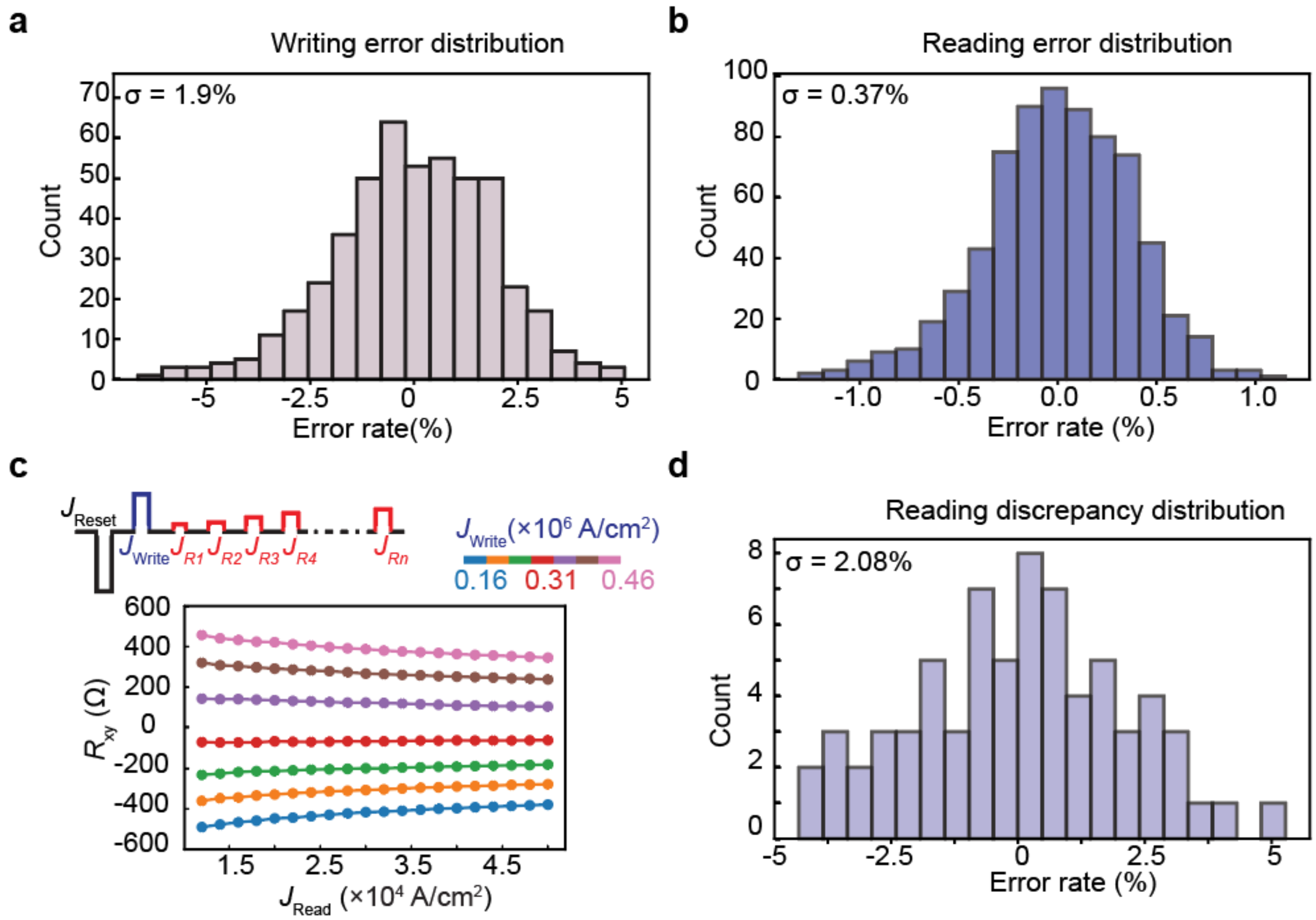

**Supplementary Figure 7**. **Memristive behavior of MTI devices D1-D4**. **a,** The write error distribution of MTI collected from statistical data of different write currents. **b,** The reading error distribution of MTI. **c,** The read discrepancy of MTI at different magnetization states using the same read current. The inset shows the test scheme, where a reset pulse ($-8.3\times10^5$ A/cm$^2$), a writing pulse, and multiple reading pulses with increasing amplitude are applied in sequence. **d,** The reading discrepancy error distribution of MTI using different reading currents. The data is collected from reading current ranging from $1.6\times10^4$ A/cm$^2$ to $3.2\times10^4$ A/cm$^2$ when the Hall resistance is between -200 Ω to 200 Ω. $\sigma$ denotes the standard deviation error.

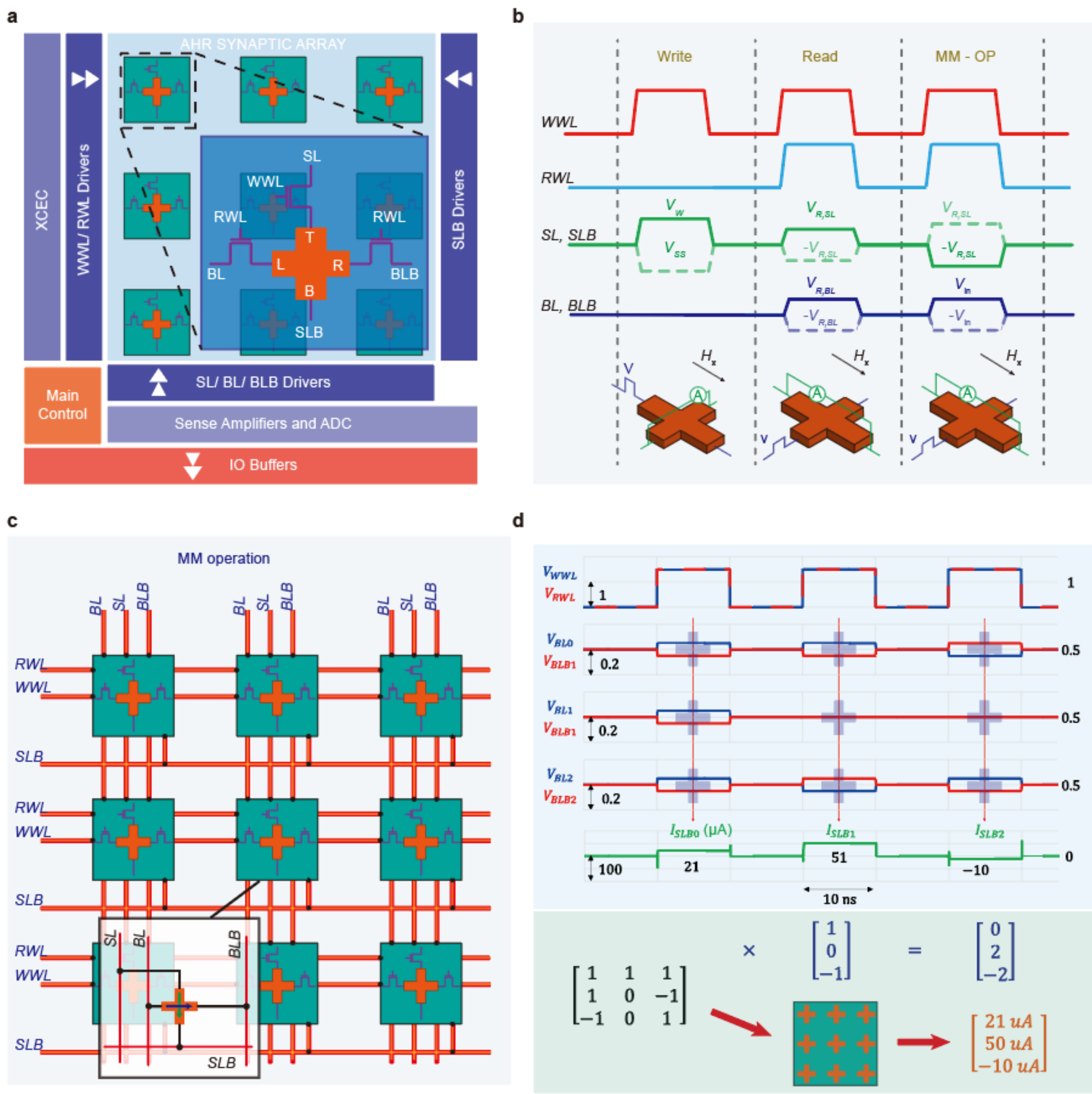

**Supplementary Figure 8**. **MTM neural network circuit design**. **a,** The schematic of the MTM neural network. The inset shows the structure of the memristor cell. **b,** The waveforms of the MTM neural network during each type of operation. **c,** The schematic description of the VMM operation. The activated bus lines are marked in red, and the inset shows the corresponding electrical connection. **d,** The circuit simulation result of a 3-input VMM operation.

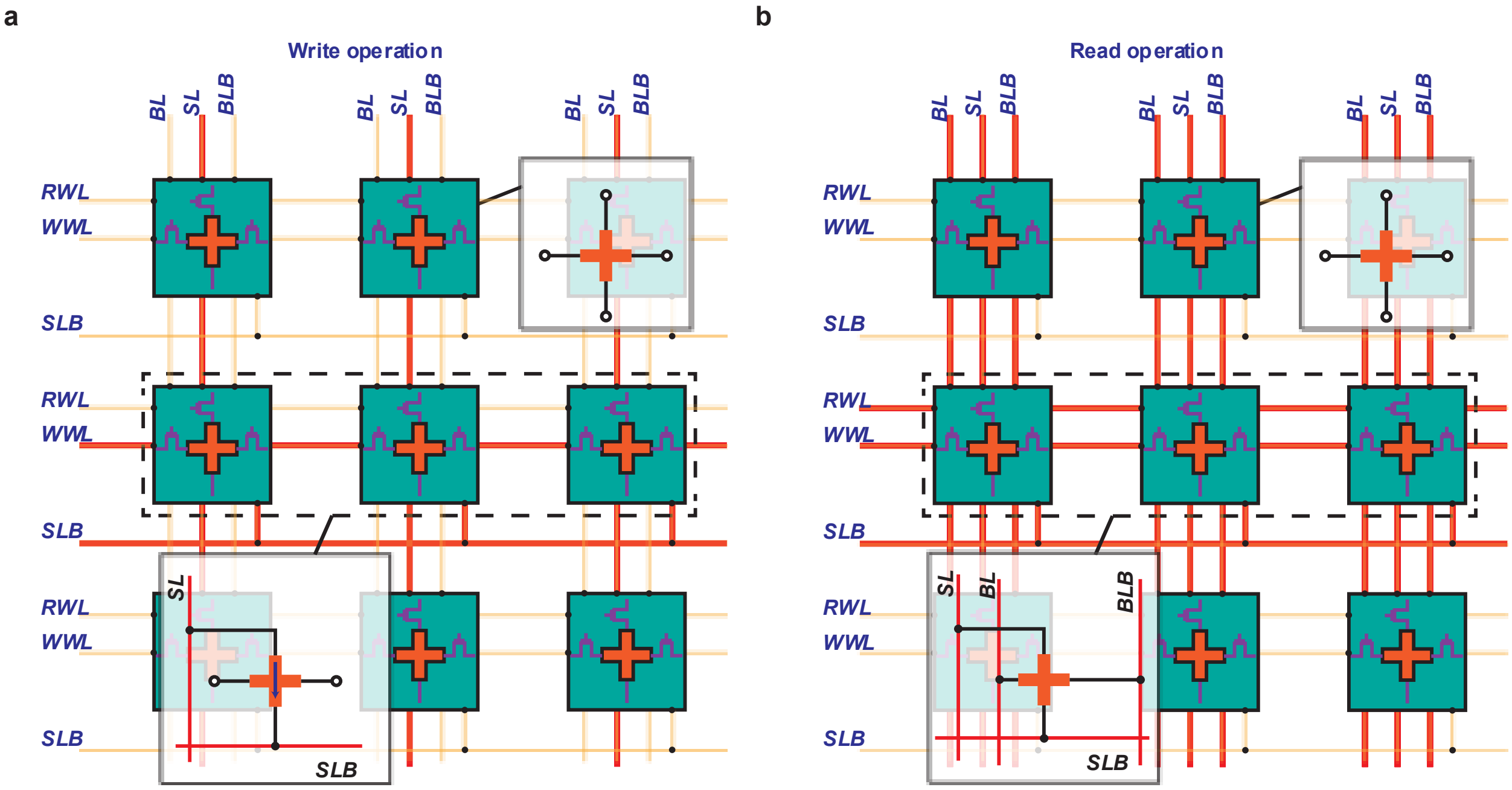

**Supplementary Figure 9**. **The schematic of write and read operations of the MTM neural network**. **a,** The schematic of the write operation. **b,** The schematic of the read operation. The inset shows the connection to each bit line of the selected and unselected devices.

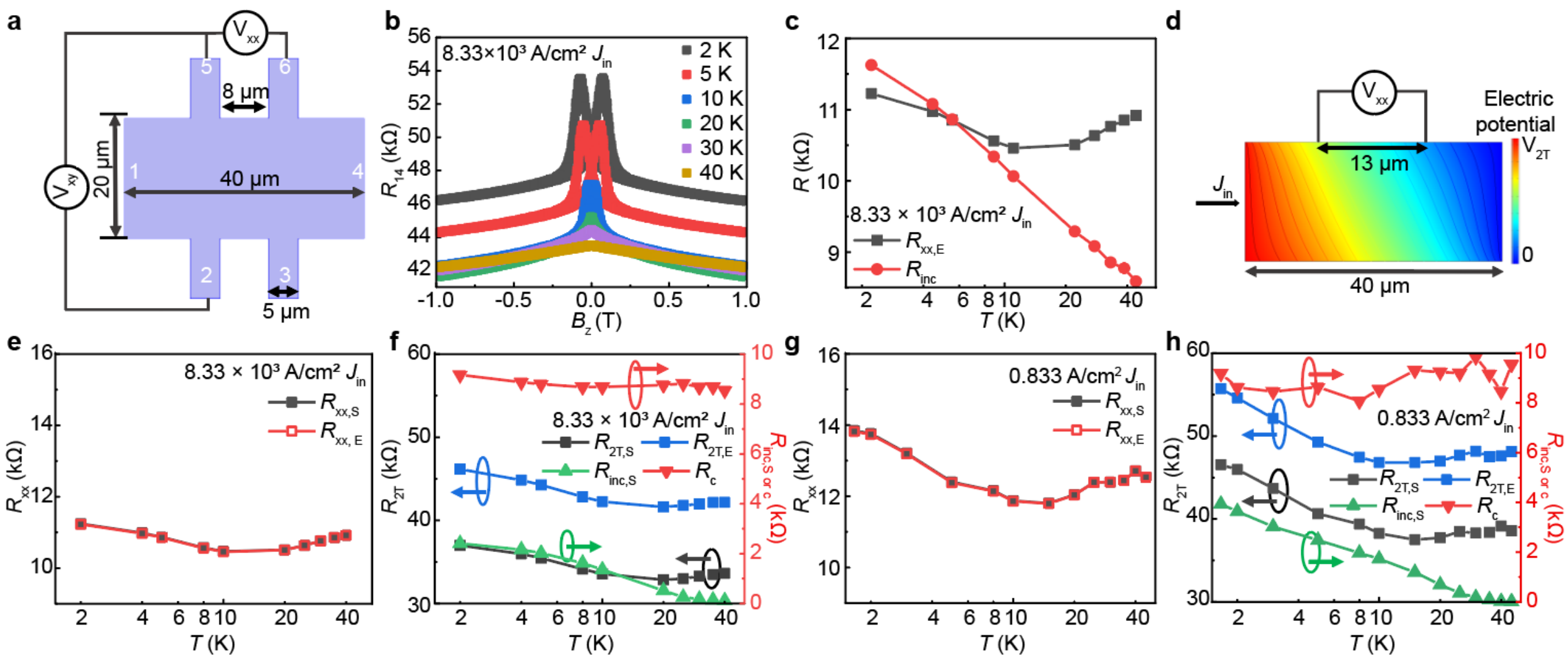

**Supplementary Figure 10**. **Two-terminal resistance of the MTI device D1**. **a,** Schematic of the Hall bar device. **b,** Out-of-plane hysteresis loops of two-terminal resistance for $R_{1,4}$. **c,** Temperature dependence of $R_{xx,E}$ and $R_{inc} = R_{2T,E} - \frac{40}{13} R_{xx,E}$ at 1T, where $R_{2T,E}$ is the experimentally measured two-terminal resistance between contacts 1 and 4. **d,** Simulation results of electric potential distribution in a rectangle sample (20 μm × 40 μm). The conductance matrix of the sample is obtained from the $R_{xx}$ and $R_{xy}$. The left and right boundaries are set to be the applied current (I) and ground, respectively. **e,** $R_{xx}$ from the experiment and simulation as a function of temperature at 1T of 10 μA ($8.33\times10^3$ A/cm$^2$) input. **f,** $R_{2T,E}$, $R_{2T,S}$, $R_{inc,S}$, $R_c$ as a function of the temperature with $R_{xx}$ from **e**, where $R_{2T,S}$ is simulated two-terminal resistance, $R_{inc,S}$ is simulated $R_{inc}$, and $R_c$ is conventional contact resistance estimated by $R_c = R_{inc} - R_{inc,S}$. The reading current is 10 uA for **e** and **f**. **g,** $R_{xx}$ from the experiment and simulation as a function of temperature at 1T of 1 nA (0.833 A/cm$^2$) input. **h,** $R_{2T,E}$, $R_{2T,S}$, $R_{inc,S}$, $R_c$ as a function of the temperature with $R_{xx}$ from **g**. The reading current is 1 nA for **g** and **h**.

**Supplementary Table 1**. Comparison of channel resistivity, SOT efficiency, and switching power metrics.

| Parameters | Ta (Control sample, 2K) | $Pt_{85}Hf_{15}$ (4K)[15] | Pt (3K)[16] | MTI (2K) |
|---|---|---|---|---|
| Channel resistivity (ρ) | 191 μΩ·cm | 25 μΩ·cm | 20 μΩ·cm | 13890 μΩ·cm |
| SOT efficiency ($\xi_{DL}$) | 0.07 | 0.2 | 0.18 | 19.2 |
| Theoretical normalized switching power ($P_{SOT} = \rho/\xi_{DL}^2$) | $3.9\times10^{-4}$ | $6.3\times10^{-6}$ | $6.2\times10^{-6}$ | $3.8\times10^{-7}$ |
| Switching current density ($J_{sw}$) | $2\times10^{7}$ A/cm$^2$ | $7.5\times10^{7}$ A/cm$^2$ | $9\times10^{7}$ A/cm$^2$ | $4.2\times10^{5}$ A/cm$^2$ |
| Experimental normalized switching power ($P_{SW} = \rho J_{sw}^2$) | $7.6\times10^{16}$ W/m$^3$ | $1.4\times10^{17}$ W/m$^3$ | $1.6\times10^{17}$ W/m$^3$ | $2.5\times10^{15}$ W/m$^3$ |

**Supplementary Table 2.** Properties of D1-D4, and D6 at 2 K. Unless specified, the reading current density is 8.3 x $10^3$ A/$cm^2$ (10 µA).

| | **$T_{turn}$ (K)*** | **$T_c$ (K)** | **$R_{xx,\square_1T}$ (kΩ/□)** | **$R_{xy_1T}$ (kΩ)** | **$R_H$ (kΩ)** | **$I_{CES}/I_{TSS}$** | **Hall angle ($R_{xy_1T}/R_{xx,\square_1T}$)** | **$J_c$** (×$10^6$ A/$cm^2$)** | **Switching ratio*** ** | **Samples applied to MTM** |
|---|---|---|---|---|---|---|---|---|---|---|
| D1 | 1.81 (1nA,9T) | 35 | 17.2 | 9.2 | 9.2 | 0.56 | 0.5 | 0.4 | 0.09 | D1-D4 |
| D2 | 1.8 (50nA,9T) | 40 | 16.1 | 8.6 | 8.6 | 0.5 | 0.5 | 0.41 | 0.1 | |
| D3 | 1.8 (50nA,9T) | 40 | 18.6 | 9.2 | 9.1 | 0.55 | 0.5 | 0.42 | 0.09 | |
| D4 | 1.94 (50nA,9T) | 40 | 18.7 | 9.5 | 9.4 | 0.58 | 0.5 | 0.42 | 0.08 | |
| D6 | 26 (1µA,1T) | 35 | 14 | 18.5 | 18 | 2.48 | 1.3 | 0.125 | 0.02 | - |

*$T_{turn}$: the turning point of $R_{xx,\square}$ in the temperature vs $R_{xx,\square}$ curve.

**$J_c$: the critical switching current density. For D1-D4, $J_c$ is defined by the current switches $R_{xy}$ to 75%. For D6, it is defined at 50%.

***Switching ratio: $R_{xy}$ switched by the current divided by the anomalous Hall resistance $R_H$.

**Supplementary Table 3**. Training parameters and final accuracy of the Floating-point (FP) software, Bipolar device (MTM), and Unipolar device.

| Task | Model | Initial lr | Weight decay | Epochs | Schedule | Test accuracy |
|---|---|---|---|---|---|---|
| MNIST | **FP** | $10^{-3}$ | $10^{-4}$ | 200 | Cosine lr | 98.27% |
| | **MTM** | $10^{-3}$ | $10^{-4}$ | 200 | Cosine lr | 98.38% |
| | **Quantized MTM** | $10^{-3}$ | $10^{-4}$ | 200 | Cosine lr | 98.21% |
| | **Unipolar** | $10^{-2}$ | $10^{-4}$ | 1000 | Cosine lr | 94.26% |
| | **Quantized Unipolar** | $10^{-2}$ | $10^{-4}$ | 1000 | Cosine lr | 93.17% |
| CIFAR | **FP** | $10^{-3}$ | $10^{-5}$ | 200 | multi step lr [80,120,160] | 91.6% |
| | **MTM** | $10^{-3}$ | $10^{-5}$ | 200 | multi step lr [80,120,160] | 91.9% |
| | **Quantized MTM** | $10^{-3}$ | $10^{-5}$ | 200 | Cosine lr | 90.9% |
| | **Unipolar** | $7\times10^{-2}$ | $2\times10^{-4}$ | 1000 | Cosine lr restart at [12,48,192] | 78.1% |
| | **Quantized Unipolar** | $10^{-3}$ | $10^{-5}$ | 1000 | Cosine lr restart at [12,48,192] | 78.3% |

**Supplementary Table 4**. Comparison of various in-memory computing solutions with different technologies and the tensor processing unit (TPU). TOPS/W are normalized following common practice, where its value is rescaled linear to the weight precision.

| | **MRAM**[14] | **Hall-FM** | **Hall-MTI** | **Hall-MTI** | **TPU**[17] |
|---|---|---|---|---|---|
| **Array Size** | 22nm*, 512×512, Device = 50nm** | | | Device = 200nm** | - |
| **Weight Availability** | Bipolar via differential | Bipolar | Bipolar | Bipolar | Bipolar |
| **$R_{on}$, $R_{off}$** | 4kΩ, 11.2kΩ | -11Ω, +11Ω | -12kΩ, +12kΩ | -12kΩ, +12kΩ | - |
| **$E_{Read}$/bit** | 0.37pJ | 0.58pJ | 0.48pJ | 0.48pJ | - |
| **$E_{Write}$/bit** | 2.72pJ | 0.86pJ | 0.38pJ | 0.48pJ | - |
| **$E_{NN_OP}$** | 3.91nJ | 7.28nJ | 1.45nJ | 1.45nJ | - |
| **TOPS/W (1b)** | 67.1 | 144 | 724 | 724 | 16 |

*CMOS 22 nm **cross section area

**Supplementary Table 5**. Parameters used in the performance evaluation.

| Circuit Level | | | | | |
|---|---|---|---|---|---|
| **Symbol** | **Description** | **Value** | | | |
| $N_{BL}$ | BL Length | 512 | | | |
| $N_{WL}$ | WL Length | 512 | | | |
| **f** | Operation Frequency | 250MHz | | | |
| **VDD** | Supply Voltage | 0.85V | | | |
| $C_{BL}$ | BL Capacitance / Cell | 0.2fF | | | |
| $C_{WL}$ | WL Capacitance / Cell | 0.4fF | | | |
| $E_{SA}$ | Sense Amplifier Energy | 4fJ | | | |
| $E_{ADC}$ | Analog-Digital Converter Energy | 1.9pJ | | | |
| **Device Level (@T=2K)** | | | | | |
| **Symbol** | **Description** | **STT MRAM** | **Hall-FM** | **Hall-MTI** | **Hall-MTI** |
| **w, l** | Device Width and Length | 50nm (diameter) | 50nm | 50nm | 200nm |
| $R_{on}$, $R_{off}$ | On, Off-State Resistance | 4kΩ, 11.2kΩ | - | - | - |
| $R_X$, $R_Y$ | X, Y-Direction Resistance | - | 850Ω, 850Ω | 31kΩ, 31kΩ | 31kΩ, 31kΩ |
| $R_{xy}$ | Hall Resistance | - | 11Ω | 12kΩ | 12kΩ |
| $V_{Write}$ | Write Voltage | 1.5V | 0.05V | 0.08V | 0.3V |
| $V_{read}$ | Read Voltage | 0.1V | 0.07V | 0.08V | 0.08V |